\documentclass{aa}

\usepackage{txfonts}
\usepackage{graphicx}

\DeclareMathOperator{\Norm}{\mathcal N}

\usepackage{natbib,twoopt}
\usepackage[breaklinks=true]{hyperref}
\bibpunct{(}{)}{;}{a}{}{,}

\makeatletter
  \newcommandtwoopt{\citeads}[3][][]{\href{http://adsabs.harvard.edu/abs/#3}%
    {\def\hyper@linkstart##1##2{}%
     \let\hyper@linkend\@empty\citealp[#1][#2]{#3}}}
  \newcommandtwoopt{\citepads}[3][][]{\href{http://adsabs.harvard.edu/abs/#3}%
    {\def\hyper@linkstart##1##2{}%
     \let\hyper@linkend\@empty\citep[#1][#2]{#3}}}
  \newcommandtwoopt{\citetads}[3][][]{\href{http://adsabs.harvard.edu/abs/#3}%
    {\def\hyper@linkstart##1##2{}%
     \let\hyper@linkend\@empty\citet[#1][#2]{#3}}}
  \newcommandtwoopt{\citeyearads}[3][][]%
    {\href{http://adsabs.harvard.edu/abs/#3}
    {\def\hyper@linkstart##1##2{}%
     \let\hyper@linkend\@empty\citeyear[#1][#2]{#3}}}
\makeatother

\begin{document}

\title{Catalog-free modeling of galaxy types in deep images}
\subtitle{Massive dimensional reduction with neural networks}
\titlerunning{Catalog-free modeling of galaxy types in deep images}
\authorrunning{Livet, Charnock, Le Borgne, de Lapparent}
\date{}
\author{F. Livet, T. Charnock\thanks{email: tom@charnock.fr}, D. Le Borgne\thanks{email: damien.le\_borgne@iap.fr}, V. de Lapparent\thanks{email: valerie.de\_lapparent@iap.fr}}
\institute{Institut d\textquotesingle Astrophysique de Paris, Sorbonne Université, CNRS, UMR 7095, 98 bis boulevard Arago, 75014 Paris, France}
    
\abstract 
{Current models of galaxy evolution are constrained by the analysis of catalogs containing the flux and size of galaxies extracted from multiband deep fields. However, these catalogs contain inevitable observational and extraction-related biases that can be highly correlated. In practice, taking all of these effects simultaneously into account is difficult, and therefore the derived models are inevitably biased as well.} 
{To address this issue, we use robust likelihood-free methods to infer luminosity function parameters, which is made possible by the massive compression of multiband images using artificial neural networks. This technique makes the use of catalogs unnecessary when observed and simulated multiband deep fields are compared and model parameters are constrained. Because of the efficient data compression, the method is not affected by the required binning of the observables inherent to the use of catalogs.}
{A forward-modeling approach generates galaxies of multiple types depending on luminosity function parameters rendered on photometric multiband deep fields that include instrumental and observational characteristics. The simulated and the observed images present the same selection effects and can therefore be properly compared. We trained a fully convolutional neural network to extract the most model-parameter-sensitive summary statistics out of these realistic simulations, shrinking the dimensionality of the summary space to the number of parameters in the model. Finally, using the trained network to compress both observed and simulated deep fields, the model parameter values were constrained through population Monte Carlo likelihood-free inference.}
{Using synthetic photometric multiband deep fields similar to previously reported CFHTLS 
and WIRDS D1/D2 deep fields and massively compressing them through the convolutional neural network, we demonstrate the robustness, accuracy, and consistency of this new catalog-free inference method. We are able to constrain the parameters of luminosity functions of different types of galaxies, and our results are fully compatible with the classic catalog-extraction approaches.}
{}

\maketitle

\section{Introduction}\label{sec:intro}

\begin{figure*}
\centering
\includegraphics[width=\textwidth]{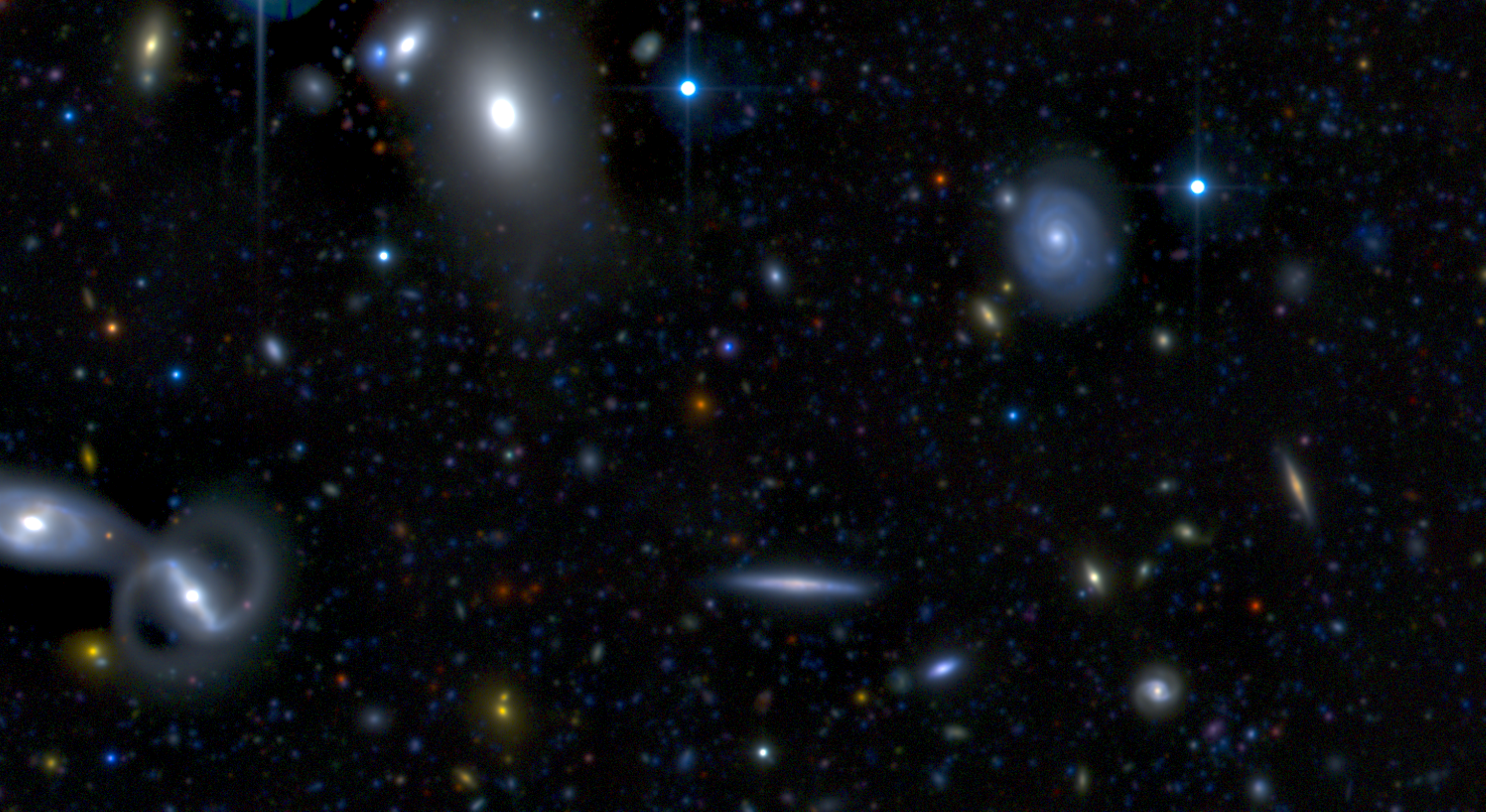}
\caption{$3\times6$ arcmin$^2$ of the 1 deg$^2$ CFHTLS D1 deep field in RGB (using the $i'$, $r'$, $g'$ filters). This shows the diversity of sizes, shapes, and colors that we can use to perform a deep statistical study of the various galaxy populations. This image reveals the complexity of the object distribution and the huge wealth of information that is available in such a field. This complexity presents a barrier in the methods we used to constrain the parameters of the luminosity functions for each population of galaxies.}
\label{fig:D1}
\end{figure*}

Traditionally, the study of galaxy evolution is based on the analysis of large sets of photometric surveys with long exposure times and a wide range of bands. These data are used with the goal of answering specific questions about galaxy evolution, such as the star-formation histories, the accretion histories, the morphological transformations, and the merging rates for galaxies of various types and masses, and the role of environment in these various characteristics.\\
\\
In a classic approach, photometric catalogs of galaxies are extracted from the images and the catalogs are used to model the joint evolution of their luminosities, colors, and sizes  with redshift. However, catalogs are subject to a variety of incompleteness effects. Rigorous studies based on flux-limited samples can be performed, but they are not optimal: in flux-limited catalogs, the very faint galaxies near the background noise are ignored by construction, which induces a Malmquist bias if this effect is not dealt with properly \citep{1922MeLuF.100....1M, 1925MeLuF.106....1M}. Moreover, some galaxies can overlap, and confusion is often a limiting effect at low luminosities \citep{1974ApJ...188..279C}, mainly at large wavelengths. Poor segmentation, morphological biases, or even the lack of a detection of galaxies can be a consequence of cosmological dimming, which decreases the surface brightness of an object by a factor $(1+z)^{-4}$, where $z$ is the redshift \citep{1934rtc..book.....T, 2014ApJ...796..102C}. Another statistical selection effect is the Eddington bias \citep{1913MNRAS..73..359E}: at a given apparent flux, the number of fainter galaxies with a flux overestimation is larger than the number of brighter galaxies with a flux underestimation. The stellar contamination of deep surveys \citep{2014MNRAS.444..846P} is another limitation of faint galaxy catalogs because some stars can be misidentified as galaxies and contaminate the results.
Finally, redshift affects the luminosities and colors of objects, and K-correction must be applied \citep{2002astro.ph.10394H, 2003ApJ...585L...5H} to correct for these effects. \\
\\
Moreover, these various selection effects are correlated in subtle ways that are very difficult to express analytically, and some may be spatially variable within a survey \citep[Appendix B]{2017MNRAS.468.2569B}. As a result, the classic approach of extracting catalogs is a complex inverse problem of trying to correct for the effects of correlated biases of these catalogs \citep{1998ASSL..231...23M,2015ApJ...801...14T}. If this is not successful, the models derived from these catalogs tend to be biased in a nontrivial way.\\
\\
In order to avoid solving the hard inverse problem of model inference from catalogs, forward modeling coupled with approximate Bayesian computation (ABC) can be used. This technique has been used in many fields and for various applications, for example, in cosmology \citep{2015JCAP...08..043A,2018JCAP...02..042K}, in modeling the initial mass function \citep{2019arXiv190411306C}, to study type Ia supernovae \citep{2013ApJ...764..116W}, and in modeling galaxy evolution \citep{2017A&A...605A...9C,2020arXiv200107727T}. In the specific case of galaxy evolution, the forward-modeling approach simulates realistic deep multiband images and compares them to observed data in order to constrain the model parameters.\\
\\
Placing constraints on the parameters of galaxy population models is difficult when catalogs  are extracted from the deep surveys that contain a huge number of galaxies with various fluxes, shapes, and sizes. Several source-extraction packages are available, for example, \texttt{SExtractor} \citep{1996A&AS..117..393B,2010ascl.soft10064B}. In a recent paper, \citet{2017A&A...605A...9C} have developed a method for binning extracted catalogs in fluxes and sizes in order to infer the parameters of the luminosity functions used in their model. They successfully measured the evolution parameters of one and two galaxy-type luminosity functions. However, the binning of the apparent magnitude distributions had to be limited to ten intervals per band (of eight bands), which nevertheless meant that the study took place in a very sparsely populated $10^8$ dimension space. Moreover, catalogs are affected by a number of extraction biases: the model point spread function (PSF), the type of magnitude used to perform the extraction (isophotal, aperture, or model), the determination of the sky background, the segmentation process, and the confusion of sources. Consequently, the analytical form for the likelihood is unknown for the parameters of the luminosity functions of the various populations of galaxies, and the information from such deep surveys could be lost by compressing it into catalogs.\\
\\
To circumvent all of these problems simultaneously, we describe here a novel method for massive data compression using neural networks in order to enable a direct comparison of observation to model in the multiwavelength images. This completely bypasses the biases induced by source extraction. We use the algorithm called information maximizing neural network (IMNN) \citep{2018PhRvD..97h3004C}, which fits a neural network that is only sensitive to the effects of the model parameters in the simulations that are obtained from our forward model. We implement this method for the first time in deep and large multiwavelength images of galaxies: the 1 deg$^2$ Canada-France-Hawaii Telescope Legacy Survey (CFHTLS) D1 deep field observed in the optical, using the MegaPrime instrument in the $u'$, $g'$, $r'$, $i'$, $z'$ filters, and in the near-infrared (IR) using the WIRCam instrument in the $J$, $H$, $Ks$ filters. We used the final releases of the TERAPIX processed data for each survey, T0007 for CFHTLS  \citep{2012yCat.2317....0H} and T0002 for WIRDS \citep{2012A&A...545A..23B}; the WIRDS images are rebinned to match the 0.186 arcsec/pixel of the optical images. A red, green, blue (RGB) image of part of the D1 deep field in the optical is shown in Fig. \ref{fig:D1}. It highlights the complexity of the data that are available for studying the various populations of galaxies. A huge wealth of information is distributed such that it is impossible to construct a likelihood that describes the probability of obtaining such a field.\\
\\
The paper is structured as follows: Sect. \ref{sec:forward} introduces the forward model we used to create simulated multiband deep fields and discusses its features. These are the luminosity function parameterization, the spectral energy distributions (SED) of a bulge+disk decomposition, internal dust extinction, stars, reddening, etc. In Sect. \ref{sec:compression} and Sect. \ref{sec:NN} we present the compression of deep fields using IMNN and discuss its characteristics. These are Fisher information, summary statistics, Gaussian and non-Gaussian likelihoods, and quasi-maximum likelihood estimators. This includes the description of the neural network architecture and the loss function we used to fit the network. Sect. \ref{sec:app1} presents a toy application with only two parameters ($\phi^*$ and $M^*$) of the luminosity functions of one spiral population. Sect. \ref{sec:app2} demonstrates a more realistic application to constrain the density parameter $\phi^*$ of both elliptical and spiral galaxies in parts of the observed CFHTLS and WIRDS D1 deep field in eight photometric bands. It also validates the methods on virtual data. As a key comparison, we use in Sect. \ref{sec:app2} the recent results of \citet{2017A&A...599A..62L}. In this paper, a $\Lambda$ cold dark matter ($\Lambda$CDM) cosmology is adopted with $\Omega_\Lambda=0.7$, $\Omega_M=0.3,$ and $H_0=70$ km s$^{-1}$ Mpc$^{-1}$.

\section{Forward model}\label{sec:forward}

In this section, we describe in detail the forward model that we used to produce mock images. These images were then used (Sect. \ref{sec:training}) to train the IMNN and also to perform the Bayesian inference in the two applications of Sects. \ref{sec:app1} and \ref{sec:app2}.

\subsection{Basis of the model}

We created mock photometric catalogs of galaxies with a code based on the \texttt{Stuff} program \citep{2010ascl.soft10067B} rewritten from C to Python by F. Livet. It now accounts for (i) distinct extinction in the bulge and disk components, and (ii) suppression of sample variance, which allows generating images with variations in the luminosity function parameters by only adding or suppressing galaxies depending on the sign of the change in the luminosity function at each absolute magnitude. This therefore results in common galaxies at the same coordinates and with identical bulge and disk parameters (see Sect. \ref{sec:app1_training}). The aim of this code is to produce realistic photometric multiwavelength catalogs of properties of galaxies from various populations. Each galaxy of these mock catalogs is decomposed as a sum of a bulge and a disk with the following properties: its apparent magnitudes in each observational band (the Sloan Digital Sky Survey (SDSS) $u'$, $g'$, $r'$, $i'$, $z'$, and the near-IR $J$, $H$, $Ks$ used for the CFHTLS and WIRDS surveys), its apparent bulge-to-total light ratio in the $g'$ band used as reference (defining its Hubble type), the characteristic sizes of its bulge and disk, its disk inclination, which is used to determine its internal extinction, and its redshift. The code uses a set of galaxy types defined by a specific luminosity function, an intrinsic bulge-to-total luminosity ratio $B/T$, the SED of the bulge and disk, the bulge and the disk scale lengths satisfying (with some dispersion) the observed scaling laws, and extinction laws. The galaxies of each type are then generated with Poisson draws in small redshift bins of $5h^{-1}$~Mpc (where $h=H_0/100$ is the reduced Hubble constant) from $z=0.001$ to $z=20$.

\subsection{Luminosity functions per galaxy type}

Each of the galaxy populations we modeled is described by its luminosity function, that is, the number of galaxies per unit comoving volume of the Universe and per magnitude interval, as a function of absolute magnitude, $M$. This probability distribution function is commonly fit by a Schechter profile \citep{1976ApJ...203..297S},
\begin{equation}\label{eq:lf}
    \Phi(M) = 0.921\phi^*\left[10^{0.4(M^*-M)}\right]^{1+\alpha}\exp\left[-10^{0.4(M^*-M)}\right]
,\end{equation}
with three parameters: $\phi^*$ the normalization density (in Mpc$^{-3}$mag$^{-1}$), $\alpha$ the faint-end slope (e.g., for $\alpha<-1$, the density of faint galaxies increases with magnitude), and $M^*$ a characteristic absolute magnitude.
The luminosity function is redshift dependent \citep{1995ApJ...455..108L}, thus two other parameters are introduced, $\phi_\text{evol}^*$ and $M_\text{evol}^*$, defined as
\begin{equation}
\label{eq:phievol}
M^* = M_0^* + M_\text{evol}^* \log(1+z)
\end{equation}

\begin{equation}     
\label{eq:mevol}
\phi^* = \phi_0^* \log(1+z)^{\phi_\text{evol}^*}
,\end{equation}
where $z$ is the redshift at which the luminosity function is measured. It has been shown that the luminosity function may evolve differently for different populations of galaxies \citep{2006A&A...455..879Z}, which means that these five parameters may be different for each galaxy population. In Sects. \ref{sec:app1} and \ref{sec:app2} we focus on constraining $M^*$ and $\phi^*$ for spirals and $\phi^*_0$ for spirals and ellipticals.

\subsection{Spectral energy distributions (SED)}

The SEDs we used to model the diversity of galaxies were sampled from the template library \citep{1980ApJS...43..393C}: "E" and "Irr". These two SEDs were assigned to the bulge and disk of a galaxy for a given morphological type in variable proportions (using $B/T$), which led to a wide range of colors for the individual galaxies: an example is shown in Fig. \ref{fig:SED}. In the current implementation, we keep the SEDs and the $B/T$ ratios fixed with redshift for all galaxy types. However, the evolution of the SEDs of individual galaxies is allowed through the evolution of the type-dependent luminosity functions and the different extinction effects (see Sect. \ref{sec:extinct}).

\subsection{Bulge component}\label{sec:bulge}

The de Vaucouleurs profile \citep{1953MNRAS.113..134D} describes how the surface brightness $I_B$ (in cd.m$^{-2}$) of a bulge or elliptical varies as a function of the apparent distance $R$ (in kpc) from the center,
\begin{equation}
    I_B(R)=I_e\exp\left[-7.669\left(\left(\frac{R}{R_e}\right)^{1/4}-1\right)\right]
,\end{equation}
where $R_e$ (in kpc) is the half-light radius or effective radius (i.e., the isophote containing half of the total luminosity of the bulge), and $I_e$  (in cd.m$^{-2}$) is the surface brightness at $R_e$. The same profile can be written in terms of magnitude as
\begin{equation}
    \mu_B(R)=M_B+8.3268\left(\frac{R}{R_e}\right)^{1/4}+5\log(R_e)+16.6337
,\end{equation}
where $\mu_B(R)$ is the bulge surface brightness (in mag.kpc$^{-2}$) at radius $R,$ and $M_B$ is the bulge absolute magnitude. It has been shown that the average effective radius follows an empirical relation with the absolute magnitude of the bulge \citep{1984AJ.....89...64B},\begin{equation}
    \langle R_e\rangle=R_0 10^{-p(M_B-M_0)}
,\end{equation}
where $R_0=1.58h^{-1}$kpc, $M_0=-20.5,$ and 
\begin{equation}p = \left\{\begin{array}{ll}
        0.3, & \text{if } M_B<M_0\\
        0.1, & \text{otherwise}
        \end{array}\right.\end{equation}
We allowed the effective bulge radius to evolve with redshift by a $(1+z)^{\gamma_B}$ factor, where $\gamma_B$ is a constant \citep{2006ApJ...650...18T, 2010ApJ...713..738W}. Furthermore, it has been shown that the intrinsic flattening, $q$, of the bulge follows a normal distribution with mean $\mu=0.65$ and standard deviation $\sigma=0.18$ \citep{1970ApJ...160..831S}. Finally, the absolute magnitude of the bulge $M_B$ is related to the total absolute magnitude of the galaxy $M$ through the relation
\begin{equation}
    M_B=M-2.5\log(B/T)
,\end{equation}
where $B/T$ is the bulge-to-total light ratio, which is assumed not to evolve for galaxies of a given type.

\subsection{Disk component}\label{sec:disk}

The exponential profile describes how the surface brightness $I_D$ (in cd.m$^{-2}$) of a disk varies as a function of the apparent distance $R$ (in kpc) from the center,
\begin{equation}
    I_D(R)=I_0\exp\left(-\frac{R}{h_D}\right)
,\end{equation}
where $h_D$ (in kpc) is the disk scale length, and $I_0$ (in cd.m$^{-2}$) is the surface brightness in the center. As for the bulge, the same profile can be written in terms of magnitude as
\begin{equation}
    \mu_D(R)=M_D+1.8222\frac{R}{h_D}+5\log(h_D)+0.8710
,\end{equation}
where $\mu_D(R)$ is the disk surface brightness (in mag.kpc$^{-2}$) at radius $R,$ and $M_D=M-2.5\log(1-B/T)$ is the disk absolute magnitude. A log-normal relation linking the disk scale length to its absolute magnitude can be fit \citep{2000ApJ...545..781D},
\begin{equation}
    h_D=h_0\exp\left[0.921\beta M_D+\Norm(0,\sigma)\right]
,\end{equation}
where $h_0=3.85h^{-1}$kpc, $\beta=-0.214,$ and $\Norm(0,\sigma)$ is a random number following a normal distribution with zero-centered mean and standard deviation $\sigma=0.36$. We allowed the disk scale-length to evolve with redshift by a $(1+z)^{\gamma_D}$ factor, where $\gamma_D$ is a constant \citep{2006ApJ...650...18T, 2010ApJ...713..738W}.

\subsection{Internal extinction}\label{sec:extinct}

The internal extinction by dust was applied separately for the bulge and disk using the interstellar extinction curve \citep{Fitzpatricketal_2007} of the Milky Way in the wavelength range from the ultraviolet (UV) to the near-IR,
\begin{equation}
    \mathrm{SED}(\lambda)=\mathrm{SED}_0(\lambda)\exp\left[-\kappa(i,\omega,\lambda)\tau(\lambda)\right]
,\end{equation}
where $\mathrm{SED}(\lambda)$ is the extincted SED of the bulge and disk, $\mathrm{SED}_0(\lambda)$ is the face-on non-extincted SED of the bulge and disk, $\tau(\lambda)$ is the extinction curve, and $\kappa(i,\omega,\lambda)$ is a coefficient depending on the wavelength $\lambda$, the inclination $i,$ and the total central opacity $\omega$ of the disk.

In order to determine the value of $\kappa(i,\omega,\lambda)$, we used the attenuation-inclination relations of \citet{2011A&A...527A.109P} to compute the differences of extinct and non-extinct magnitudes for the bulge and the disk. Then, we interpolated these relations to obtain the values of $\kappa(i,\omega,\lambda)$ for a random inclination $i$ between $0$ and $\pi/2$ rad. To determine the total central opacity $\omega$ of the galaxy, we used the values obtained from a Markov chain Monte Carlo (MCMC) analysis \citep{2013MNRAS.432.2061C} of the SDSS Data Release 7 and of \citet{2009ApJS..182..543A},\begin{equation}\omega=\left\{\begin{array}{ll}
        1.90_{-0.10}^{+0.23} & \text{for bulgeless galaxies}\\\\
        1.68_{-0.10}^{+0.19} & \text{for galaxies with bulges.}
        \end{array}\right.\end{equation}
In our applications in Sects. \ref{sec:app1} and \ref{sec:app2}, we only use $\omega=1.68_{-0.10}^{+0.19}$ because we consider elliptical and/or spiral populations of galaxies, and we model both populations with a bulge ($B/T=1$ for the elliptical and $B/T=0.2$ for the spiral galaxies). Finally, in each redshift bin, we extincted the magnitude of the bulge and of the disk for the effect of the intergalactic medium on a source \citep{1996MNRAS.283.1388M}.

\begin{figure}
\includegraphics[width=\columnwidth]{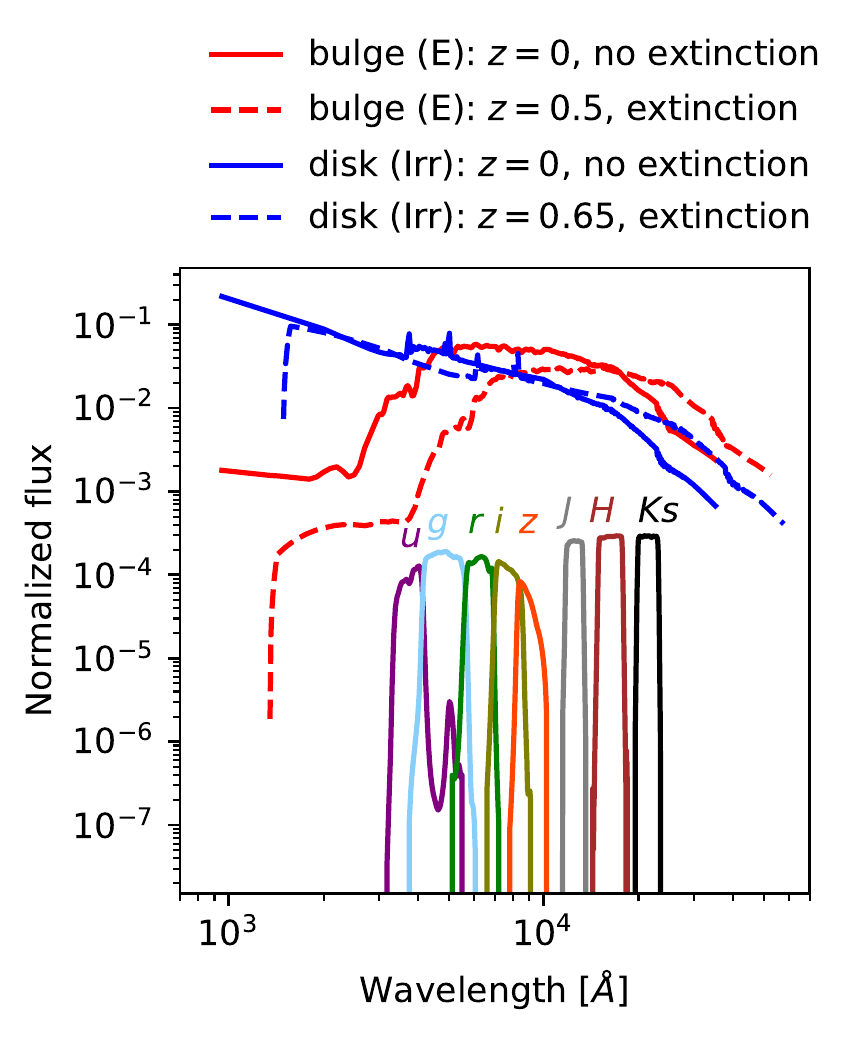}
\caption{Example of the evolution of the bulge and disk SEDs from far-UV to near-IR. The plain red curve shows the generic E template SED of an elliptical galaxy of \citet{1980ApJS...43..393C} at redshift $z=0$ and without extinction. The dashed red curve shows the same SED at redshift $z=0.5$ and with extinction (where $\omega=1.68$ and $i=45^\circ$, see Sect. \ref{sec:extinct}).
The plain blue curve shows the generic Irr template SED of \citet{1980ApJS...43..393C} used here for the disk modeling of spirals at redshift $z=0$ and without extinction. The dashed blue curve shows the same SED at redshift $z=0.65$ and with extinction (where $\omega=1.68$ and $i=45^\circ$).\newline
This shows that even though these SEDs do not evolve explicitly, the evolution of the type-dependent luminosity functions and the different extinction effects implicitly allow for an SED evolution of individual galaxies.
The plain curves are normalized so that the integral of the template SED multiplied by the SED of the reference SDSS $g'$ band equals 1. The eight filters used in this paper ($u'$, $g'$, $r'$, $i'$, $z'$, $J$, $H$, $Ks$) are shown at the bottom, and their responses are multiplied by a factor $0.0003$.}
\label{fig:SED}
\end{figure}

\subsection{Stars}\label{sec:stars}

\begin{figure}
    \includegraphics[width=\columnwidth]{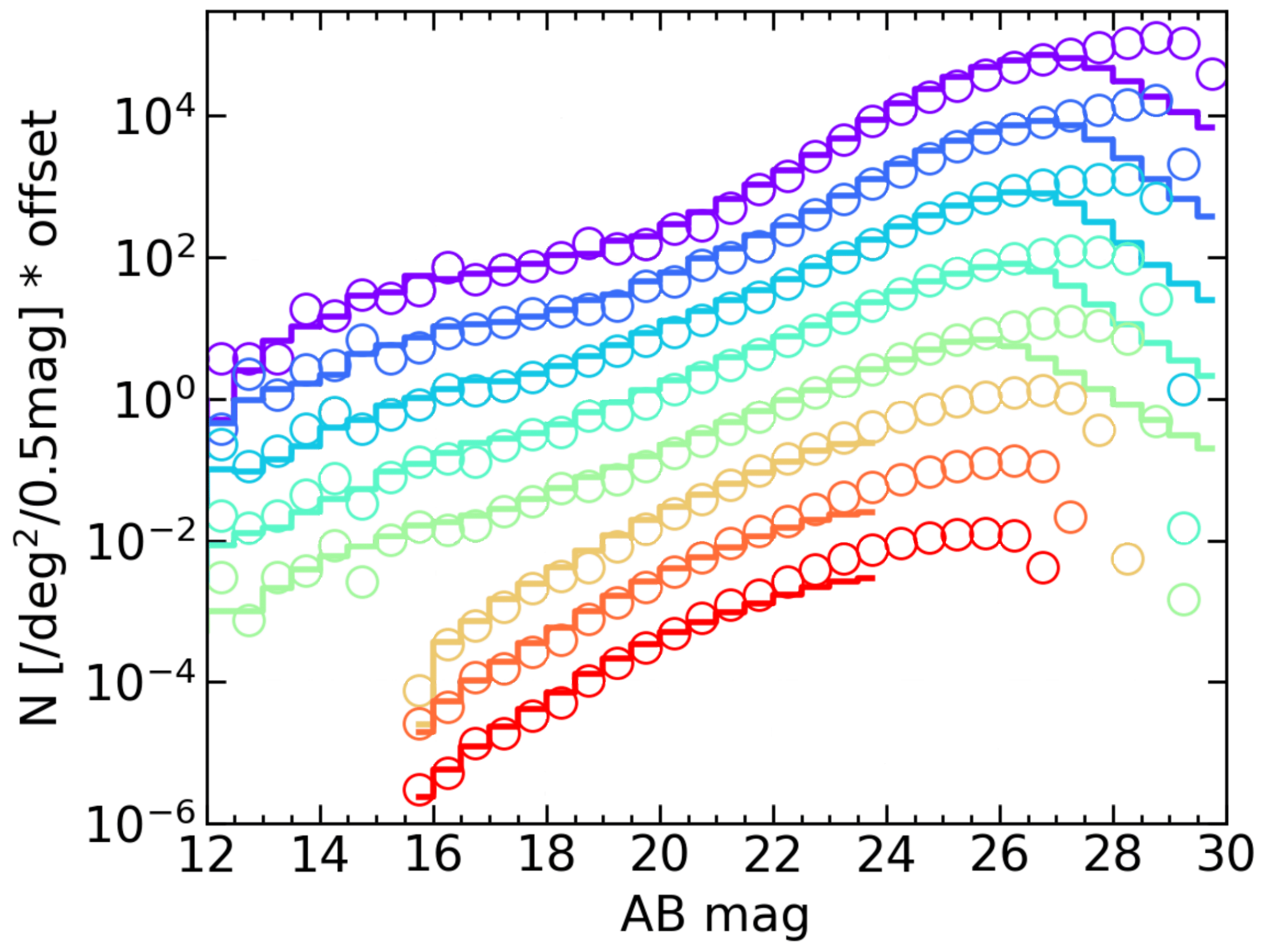}
    \caption{Differential source counts in $u'$, $g'$, $r'$, $i'$, $z'$, $J$, $H$, $Ks$ (from top to bottom). Histograms show CFHTLS observations: star+galaxy counts built from $u'$, $g'$, $r'$, $i'$, $z'$ source catalogs from MegaPrime D1+D2 \citep{2012yCat.2317....0H}, and $J$, $H$, $Ks$ galaxy counts taken from WIRCam D1+D2+D3+D4 \citep{2012A&A...545A..23B}. Our fiducial model is shown as empty circles (without stars for near-IR bands, like the observations). For clarity, the counts in each band are regularly offset vertically downward by 1 dex from $u'$ to $Ks$. This graph shows that the magnitudes of the galaxies in the forward model agree well with the observations down to their completeness limits in all eight photometric bands.}
    \label{fig:counts_all}
\end{figure}

\begin{figure}
    \includegraphics[width=\columnwidth]{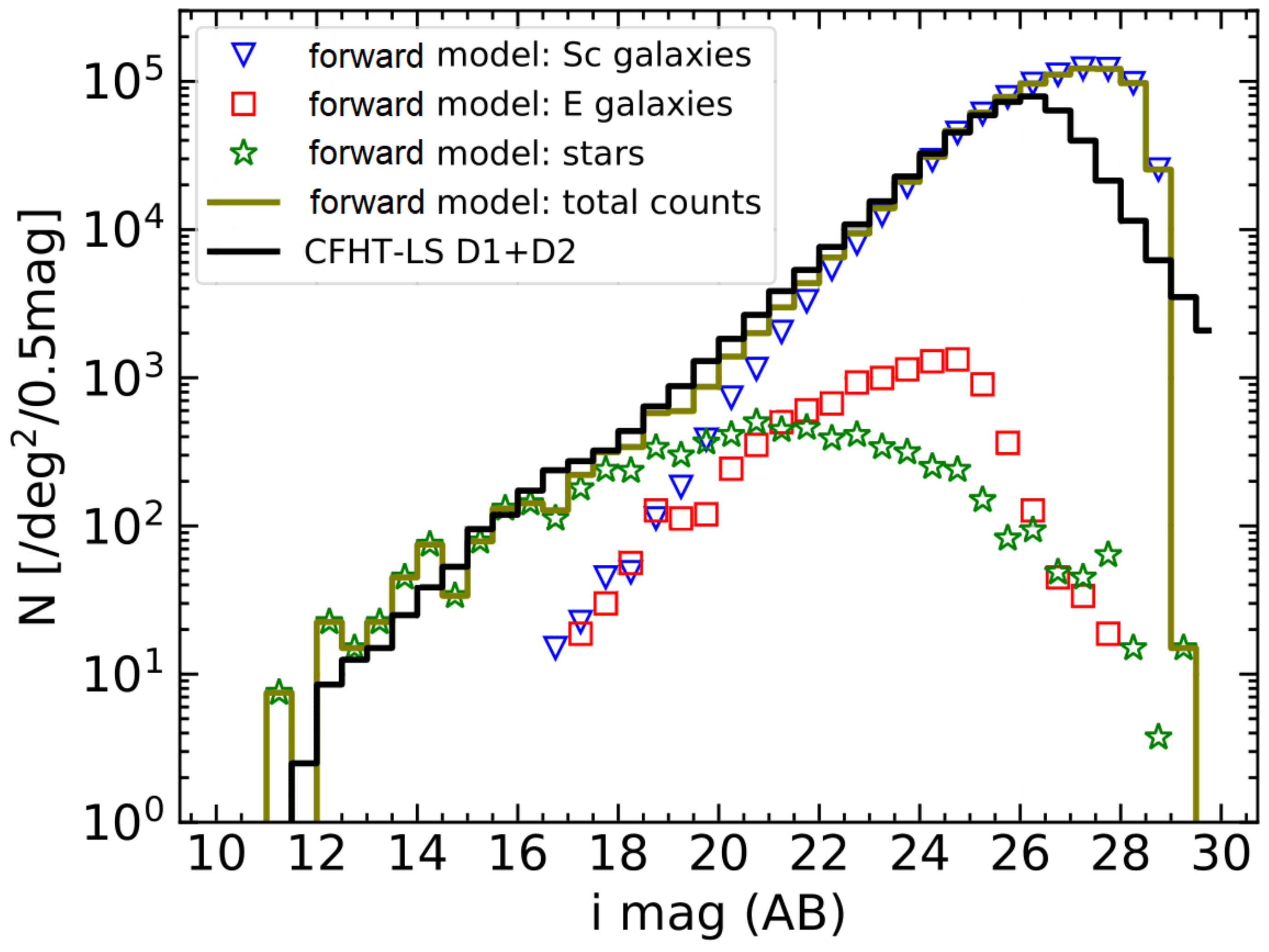}
    \caption{Differential counts (stars and galaxies) in the $i'$ band for the CFHTLS D1+D2 fields matching our fiducial model decomposed into stars (Besançon model), elliptical galaxies, and spiral galaxies down to the completeness limit. This graph shows the dominance of stars and spiral galaxies at the bright and faint end, respectively, and the difficulty of constraining the modeling of elliptical galaxies because there are so few of them. At the faint end, the completeness of the CFHTLS source extractions is limited to $i \simeq 26,$ whereas the fiducial model shows the fainter input source distribution.}
    \label{fig:counts}
\end{figure}

In order to obtain realistic deep fields, we added stars using the Besançon model \citep{2003A&A...409..523R} web service, which includes their recent improvements \citep{2012A&A...538A.106R, 2014A&A...569A..13R,  2015A&A...581A.123B, 2017A&A...602A..67A}. This model was run at the coordinates of the CFHTLS D1 and D2 deep fields in the eight photometric bands. When a smaller image than the 1 deg$^2$ deep field was used, a Poisson random draw was performed to select only the relative number of stars in the subarea. Fig. \ref{fig:counts_all} shows the total differential counts for the D1+D2 deep fields and our forward model including stars in all bands, and Fig. \ref{fig:counts} the decomposition into stars and both types of galaxies in the $i'$ band alone.
The similarity of the model and the observations is clear in the total density of objects. The stars represent only a few percent of the objects below the completeness limit of $g \simeq 26.5$ mag.

\subsection{Milky Way reddening}

In order to correct the apparent magnitudes of the galaxies for the effect of dust in the Milky Way at the specific location of the CFHTLS D1 deep field, we corrected each observed magnitude with the parameterization of \citet{1990ApJS...72..163F} in the UV and spline fitting in the optical and IR with the $R_V$ factor of 3.1 and an average $E(B-V)$ of 0.02 using the Python \texttt{extinction} package\footnote{https://github.com/kbarbary/extinction}. These correction values are given in Table \ref{table:skymaker}.

\subsection{Image generation}

The catalogs of galaxy magnitudes and sizes generated following the procedure described above were converted into images using \texttt{SkyMaker} \citep{2009MmSAI..80..422B}. \texttt{SkyMaker} renders the modeled galaxies on a high-resolution pixel grid and convolves them with an auto-computed realistic PSF \citep{2011ASPC..442..435B}. The PSF is not completely suitable for very bright stars or galaxies that are above the saturation level, which explains why the simulations do not show diffraction patterns. In addition, for each simulation, we used the weight maps provided by \citet{2012yCat.2317....0H} and \citet{2012A&A...545A..23B} to realistically model the dead pixels and the differences between the charge-coupled device (CCD) detectors. The model was then subsampled at the final image resolution and centered at some specified coordinates. \texttt{SkyMaker} finally adds the sky background, saturation, photon (Poisson) noise and read-out (Gaussian) noise. The various parameters we used for each passband are shown in Table \ref{table:skymaker} (the long exposure times of the CFHTLS deep fields are reflected in the high effective gains with a 1-second exposure time).

\subsection{Data conditioning}

Fluxes from galaxies and/or stars may have a wide dynamic range that reaches the saturation level in the observed and simulated deep-field images. This can be problematic when a neural network is fit on these fluxes because it will have more effect on very high input values than it does on the bulk. Therefore we applied the following inverse hyperbolic sine transform to each pixel of the images:
\begin{equation}
    g(x)=\text{sinh}^{-1}(x)=\log\left(x+\sqrt{x^2+1}\right)
.\end{equation}
This function reduces the dynamic range of the fluxes and allows successful fitting of the neural network with the observed and simulated deep-field images, see Sect. \ref{sec:compression}.

\section{Compression of deep fields}\label{sec:compression}

In this section, $D$ is a set of $n$ independent observations or simulations (in the form of multiband deep-field images where all pixels have been assembled into a single long vector), $d_i$ ($i=1,\dots,n$) is one specific observation/simulation of $D$ (i.e. a vector with millions of components), and $\theta$ is a vector with $p$ components $\theta_\alpha$ ($\alpha=1,\dots,p$).

\subsection{Compression through neural networks}

In the past two decades, progress in the field of pattern recognition or classification on images has been tremendous. It was shown in 2004 that standard neural networks can be greatly accelerated by using graphics processing units (GPUs), whose   implementation is 20 times faster than the same implementation on central processing units \citep[CPUs; ][]{OH20041311}. Convolutional neural networks were shown to outperform more traditional machine-learning methods for images \citep{CIRESAN2010}: a deep convolutional neural network won the ImageNet Large Scale Visual Recognition Challenge in 2012 \citep{NIPS2012_4824}.\\
\\
To compress the images, we used a neural network whose parameters were fit via the maximization of the logarithm of the determinant of the Fisher information matrix \citep{2018PhRvD..97h3004C}. Because of the multiscale properties and the various shapes of the galaxies, we decided to use a fully convolutional inception architecture; see Sect. \ref{sec:NN} for a full description.

\subsection{Fisher information}

The Fisher information is the amount of information on the unknown model parameters that the set of observations or simulations, $D$, contains when considered from a particular position $\theta$ of the parameters space.\newline
The probability of the data given the model is described by a likelihood function, $\mathcal{L}$, whose value, $\mathcal{L}(D|\theta)$, describes how likely the observations or simulations $D$ are, given particular values for the model parameters, $\theta$.\newline
We define the score of a set of observations, $D$, with likelihood function $\mathcal{L}$ depending on the parameters $\theta$,
\begin{equation}
    S\left(D,\theta\right)=\nabla_\theta\log \mathcal{L}(D|\theta)
.\end{equation}
The score indicates the sensitivity of the likelihood function to small changes in the parameters of the model. The variance of the score is the Fisher information matrix at some value of the parameters, $\theta$,

\begin{equation}\label{eq.fisher}
    F(D,\theta)=\mathbb{E}\left[S\left(D,\theta'\right)S\left(D,\theta'\right)^T\bigg\rvert\theta'=\theta\right]
.\end{equation}

\subsection{Gaussian likelihood function}\label{sec.gaussian_like}

If we assume that the data set $D$ for parameters $\theta$ satisfies a Gaussian likelihood function, $\forall i\in\{1,\dots,n\}$,
\begin{equation}\label{eq.gaussian}
        -2\log\mathcal{L}(d_i|\theta)=\left(d_i-\mu\right)^TC^{-1}\left(d_i-\mu\right)+\log (2\pi\det C)
,\end{equation}
where $\mathcal{L}(d_i|\theta)$ is the likelihood function describing the probability of a single data observation, $d_i$, given parameters $\theta$, $\mu=\mu(\theta)=\frac{1}{n}\sum_{k=1}^{n}d_k$ is the mean vector over all the data depending on $\theta$ and $C=\frac{1}{n-1}\sum_{k=1}^{n}\left(d_k-\mu\right)\left(d_k-\mu\right)^T$ is the covariance matrix of the data. Assuming the covariance to be independent of the parameters $\theta$ (see Sect. \ref{sec:loss} for a full explanation) and replacing the Gaussian likelihood of Eq. \eqref{eq.gaussian} in Eq. \eqref{eq.fisher}  for the data set, $D$, yields the Fisher information matrix,
\begin{equation}
    F(D,\theta)=\nabla_\theta\mu^TC^{-1}\nabla_\theta\mu\label{eq:FG}
.\end{equation}
By maximizing the likelihood function with respect to the parameters, that is, finding the parameters $\widetilde{\theta}$ at which the score is zero, we have a quasi-maximum likelihood estimator for any unknown observation or simulation $d$,
\begin{equation}\label{eq:MLE}
    \widetilde{\theta}=\theta + F^{-1}(D, \theta)S(d, \theta),\end{equation}
where here, $S(d,\theta)=\left(\nabla_\theta\mu\right)^TC^{-1}\left(d-\mu\right)$.

\subsection{General case: Unknown likelihood}\label{sec:general-case}

For a problem such as infering model parameters from deep galactic fields, the likelihood is not Gaussian, and worse, is not known. Developing the idea of score compression \citep{2018MNRAS.476L..60A} and the MOPED algorithm \citep{2000MNRAS.317..965H} in parallel, the proposed method \citep{2018PhRvD..97h3004C} used here considers a neural network, $f : D\rightarrow T$, which compresses the data set, $D=(d_1,\dots,d_n)$, to a set of summary statistics, $T=(t_1,\dots,t_n)$, where each $t_i$ is a vector of $p$ components, under the constraint of maximizing the Fisher information. Each piece of data, $d_i$, can be very large (e.g., in our application, each simulation is a large multiband deep-field image with $1024\times 1024$ pixels) but can be compressed, in practice without any loss, down to a vector of $p$ components, the number of model parameters ($p=2$ in the applications of Sects. \ref{sec:app1} and \ref{sec:app2}).\\
\\
Because each piece of data, $d_i$, is described in our model by the parameters $\theta$, the corresponding summary statistic, $t_i$, is also dependent on $\theta$. The neural network $f$ can be seen as a function that transforms the original unknown likelihood function $\mathcal{L}(d_i|\theta)$ of the data into an asymptotically Gaussian likelihood function $\mathcal{L}(t_i|\theta)$ of the summary statistics, $\forall i\in\{1,\dots,n\}$,

\begin{equation}
    -2\log\mathcal{L}(t_i|\theta)=\left(t_i-\mu_f\right)^TC_f^{-1}\left(t_i-\mu_f\right)+\log (2\pi\det C_f)
,\end{equation}
\newline
where, with the same formalism as the Sect. \ref{sec.gaussian_like}, $f(d_i)=t_i$, $\mathcal{L}(t_i|\theta)$ is the likelihood function for a single summary statistic $t_i$ obtained from the data $d_i$ via the transformation $f$ given parameters $\theta$, $\mu_f$ is the mean vector of the summary statistics and $C_f$ is the covariance matrix of the summary statistics. Consequently, the Fisher information matrix, ignoring covariance dependence on the parameters, is
\begin{equation}
    F=\nabla_\theta\mu_f^TC_f^{-1}\nabla_\theta\mu_f
    \label{eq:IMNNFG}
.\end{equation}
As shown in \citet{2018MNRAS.476L..60A}, and following the MOPED algorithm \citep{2000MNRAS.317..965H}, a compression scheme can be designed to be optimal in terms of lossless data compression for data where the likelihood function is known. To do so, we note that the information inequality states that the variance of some observable statistic, $t_i(\theta)$, of parameters $\theta$, is bounded by
\begin{align}
    \mathbb{E}\left[t_i(\theta')\right. & \left.t_i(\theta')^T|\theta'=\theta\right] \ge\nonumber\\
    &\nabla_\theta\mathbb{E}\left[t_i(\theta')^T|\theta'=\theta\right]F^{-1}\nabla_\theta\mathbb{E}\left[t_i(\theta')|\theta'=\theta\right].
    \label{eq:infoinequality}
\end{align}
By equating the summary statistic with the score, $t_i(\theta)=S(d_i, \theta),$ we note that the left-hand side of Eq. \eqref{eq:infoinequality} is equivalent to the Fisher information defined in Eq. \eqref{eq.fisher}. Because the gradient with respect to the parameters commutes with the expectation value,
\begin{align}
    \nabla_\theta\mathbb{E}\left[t_i(\theta')|\theta'=\theta\right] & = \mathbb{E}\left[\nabla^2_{\theta'}\log\mathcal{L}(D|\theta')|\theta'=\theta\right]\nonumber\\
    &=-F(D, \theta).
\end{align}
Substituting this back into Eq. \eqref{eq:infoinequality} shows that using the score function as a summary statistic saturates the information inequality \citep{LehmCase98} and therefore constitutes an optimal, lossless compression. This transformation yields the same quasi-maximum likelihood estimator $\widetilde{\theta}$ as Eq. \eqref{eq:MLE}, but for the summary statistics $t$ of an unknown observation or simulation $d$,
\begin{equation}\label{eq:MLE_summaries}
    \widetilde{\theta}=\theta + F^{-1}(D, \theta)S(t, \theta),\end{equation}
where here, $S(t,\theta)=\left(\nabla_\theta\mu_f(\theta)\right)^TC_f^{-1}\left(t-\mu_f(\theta)\right)$.\\
\\
By fitting the parameters of a neural network to maximize the logarithm of the determinant of Eq. \eqref{eq:IMNNFG}, we can therefore approximate the score compression for general, nonlinear likelihoods and thereby efficiently compress our large multiband images to approximately sufficient statistic vectors.

\section{Training the inception network}\label{sec:NN}

\subsection{Inception network}

\begin{figure}
    \includegraphics[scale=0.7]{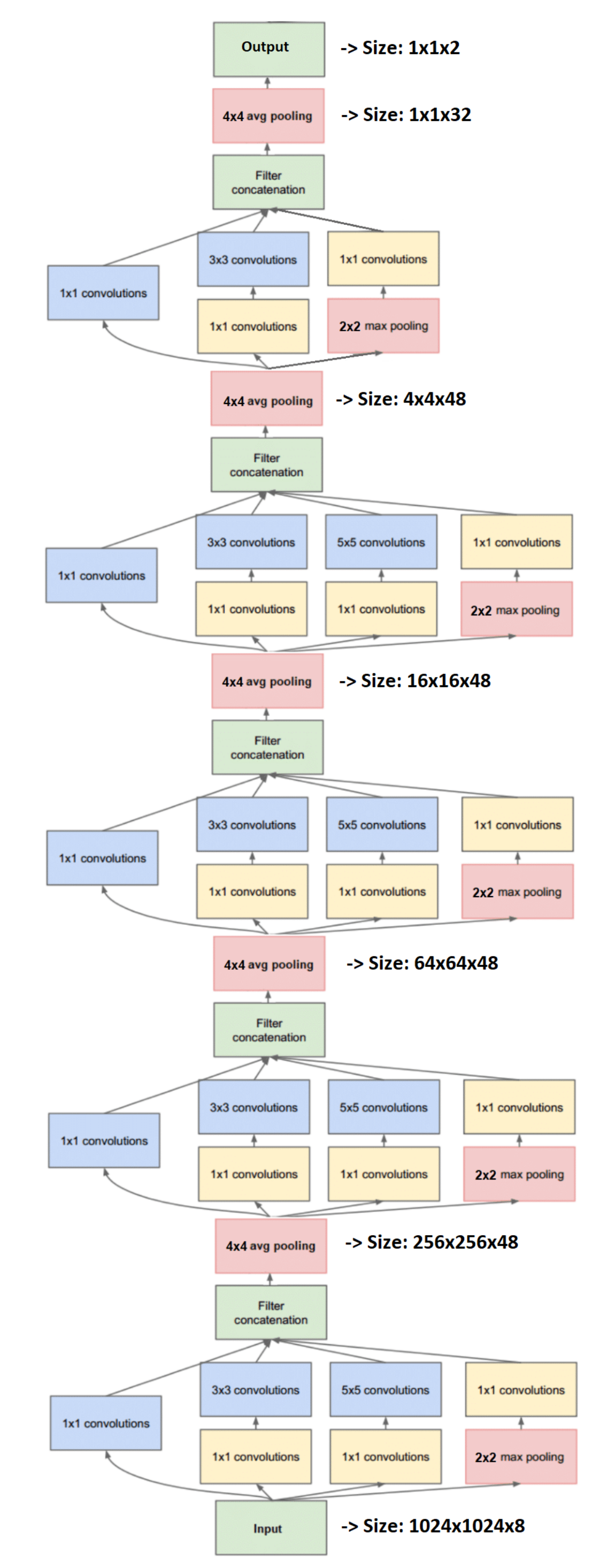}
    \caption{Fully convolutional inception network to perform the compression, see Table \ref{table:architecture} for a full description. Each inception block is composed of parallelized convolutions that simultaneously process the same input at different scales to extract different features, then concatenates the full output. After each inception block, the input is compressed with a $4\times 4$ pooling layer to decrease the resolution by a factor 4, then we indicate the current size. Finally, the output layer allows a compression down to the number of parameters of the model and is the summary statistics vector of Sect. \ref{sec:general-case}.}
    \label{fig:architecture}
\end{figure}

Galaxies can be of different sizes and shapes, hence salient parts in the image can have extremely large variation in size. Because of this huge variation in the distribution in scale of the information, choosing the correct kernel size for the convolution operation is delicate but critical. Moreover, very deep networks (several layers with many neurons) are prone to overfitting, and naively stacking large convolution operations is computationally expensive. A way to circumvent these problems was developed \citep{Szegedy_2015_CVPR} using a so-called inception block: a parallel mixture of convolutions and pooling layers (see an example in Fig. \ref{fig:architecture} with a series of five consecutive inception blocks). Inception blocks allow studying the input data at different scales in parallel and to extract features of different sizes from the same input. A succession of inception blocks can be used to create a full network architecture, which is called an inception network. The network architecture we used in the applications of Sects. \ref{sec:app1} and \ref{sec:app2} is shown in Fig. \ref{fig:architecture}.\\
\\
Because distant galaxies are generally spread over only a few pixels in size, we used kernels with pixel size $1\times 1$, $3\times 3$ (equivalent to) and $5\times 5$ (equivalent to) in each of the layers\footnote{The indicated $3\times3$ kernels in Fig. \ref{fig:architecture}  correspond to $1\times3$ and $3\times1$ kernels performed in parallel and then concatenated (this reduces the number weights, but is strictly equivalent, see \citealt{2015arXiv151200567S}), and similarly for the $5\times5$ kernels.}. As part of the inception architecture, $1\times 1$ convolutions were performed before each $3\times 3$ or $5\times 5$ kernel sized convolutions to reduce the depth, hence the number of kernels, of each block. Each convolution had eight channels (i.e., depths), which kept the total number of parameters relatively low (12 800 parameters in our applications) compared to traditional inception networks (with millions of parameters). At the end of each inception block, we concatenated (in depth) each output of the convolutions and then applied an average pooling using the mean over $4\times 4$ patches of the output to decrease the resolution of the image by $4$. This process was applied several times until the output was of the expected dimension, that is, the number of parameters of the model. Table \ref{table:architecture} gives a complete description of each component of the inception network. Specifically, the full $1024\times 1024\times 8$ input data used in the applications described in Sects. \ref{sec:app1} and \ref{sec:app2} were massively compressed by the network down to the number of parameters in the corresponding luminosity function model. These compressed summaries were then used to compute the quantities defined in Sect. \ref{sec:general-case}: $\mu$, $C,$ and finally, $F$. Our inception network is fully convolutional, we do not use any fully connected layer so that complete translational invariance is incorporated.

\subsection{Loss function}\label{sec:loss}

Normally, the loss function is a measure of the fit of the neural network to the data. However, with the IMNN, the loss function is a measure of the amount of information that is extracted from the data. We used IMNN to maximize the Fisher information of the output summaries obtained from the network. As a consequence, during the training, we measured the determinant of the Fisher matrix, which we wished to maximize. However, because the Fisher information is invariant under linear scaling of the summaries, the magnitude of the summaries needs to be controlled, which can be done by constraining the covariance matrix (see below). We therefore defined two loss functions,
\begin{equation}
    \Lambda_F = -\log\left[\det F\right]
    \label{eq:fisher}
\end{equation}
\begin{equation}
    \Lambda_C = 0.5\left(||C_f-I||_\text{F}^2+||C_f^{-1}-I||_\text{F}^2\right)
    \label{eq:covar}
,\end{equation}
where $||.||_\text{F}$ is the Frobenius norm. The first loss function $\Lambda_F$ measures the Fisher information, and the second loss function $\Lambda_C$ measures the difference of the covariance from the identity matrix. We tried to find a network parameter configuration such that both loss functions are minimal. Minimizing $\Lambda_F$ maximizes the information, while minimizing $\Lambda_C$ provides a covariance of the summaries that is close to the identity matrix. Keeping the covariance relatively close to identity fixes the scale of the summary statistics while providing a covariance that is mostly independent of the parameters, which justifies dropping the covariance term in Eqs. \eqref{eq:FG} and \eqref{eq:IMNNFG}. Several techniques \citep{rug01:000011811} exist to achieve the minimization process of such a multiobjective problem (e.g., linear scalarization and $\epsilon$-constraint). We used a continuously penalized optimization in which we combined the loss functions $\Lambda_F$ and $\Lambda_C$ as follows:
\begin{equation}
    \Lambda=\Lambda_F+r_{\Lambda_C}\Lambda_C
,\end{equation}
where 
\begin{equation}
r_{\Lambda_C}=\frac{\lambda\Lambda_C}{\Lambda_C+\exp(-\alpha\Lambda_C)}
\end{equation}
is a sigmoid function with user-defined parameters $\lambda$ and $\alpha$. As a result, when the covariance is far from identity, the $r_{\Lambda_C}$ function is large and the optimization concentrates on bringing the covariance and its inverse back to identity.

\subsection{Training of the network}\label{sec:training}

\begin{figure*}
    \centering
    \includegraphics[width=\textwidth]{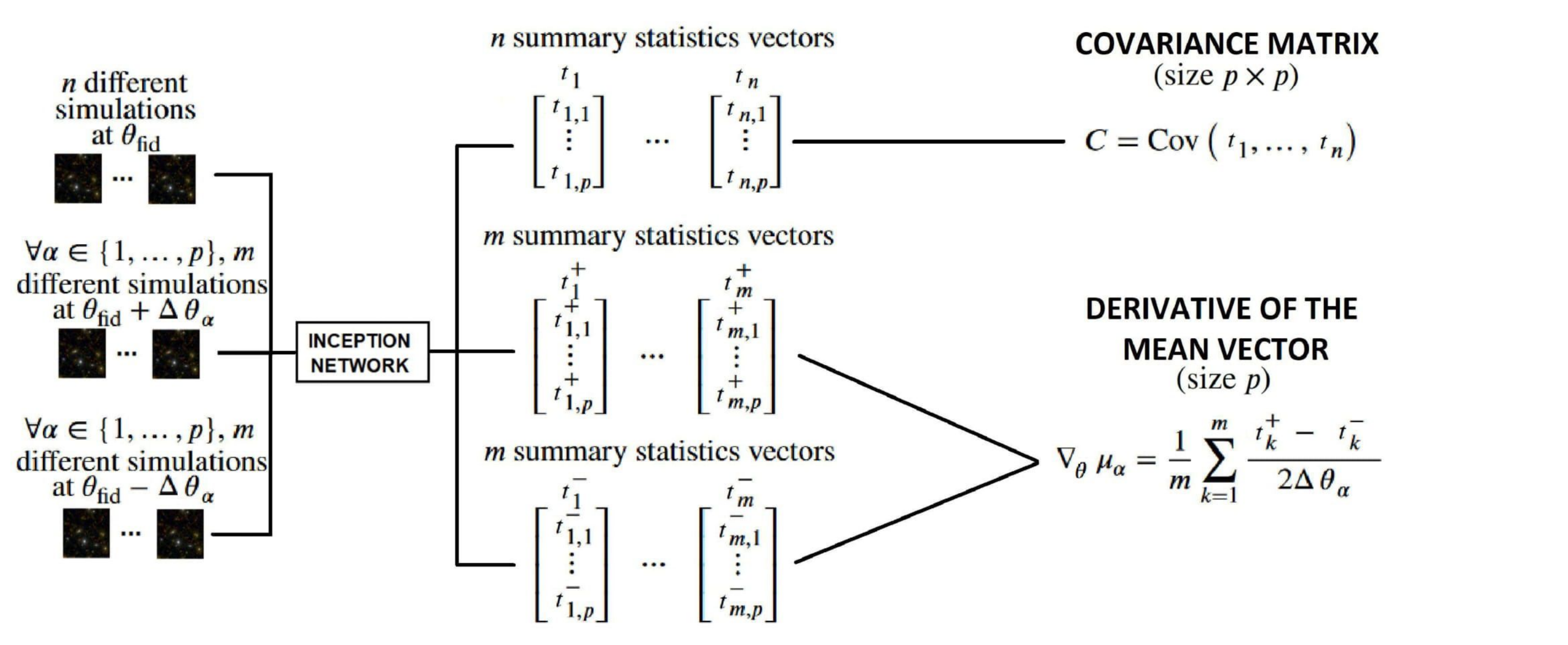}
    \caption{Schematic description of the training of the network to obtain the covariance matrix and the derivative of the mean vector that are used in Eq. \eqref{eq:IMNNFG} to obtain the Fisher information matrix. The top, middle, and bottom lists of images form what we call the training set and are generated with fiducial parameters $\theta_\text{fid}$ (with $\theta_\text{fid} + \Delta\theta_\alpha$ and $\theta_\text{fid} - \Delta\theta_\alpha$, for each $\alpha\in\{1,\dots,p\}$). From the first collection of images, we compute the covariance matrix, and from the second and third collections of images, we compute the derivative of the mean vector for each $\alpha\in\{1,\dots,p\}$.}
    \label{fig:training}
\end{figure*}

Following the description of the IMNN code\footnote{https://bitbucket.org/tomcharnock/imnn} \citep{2018PhRvD..97h3004C} to train the network when exact gradients of the simulator are not available, we need a vector of fiducial parameter values, $\theta_\text{fid}$, and a vector of deviation values about the fiducial, $\Delta\theta$.
From these values, we generate a training set of images consisting of a first collection of $n$ images generated with values at the fiducial parameters $\theta_\text{fid}$, a second collection of $m$ images generated with values $\theta_\text{fid} + \Delta\theta_\alpha$\footnote{$\theta_\text{fid} + \Delta\theta_\alpha=\left(\theta_{\text{fid},1},\dots,\theta_{\text{fid},\alpha}+ \Delta\theta_\alpha,\dots,\theta_{\text{fid},p}\right)$}, for each $\alpha\in\{1,\dots,p\}$ and a third collection of $m$ images generated with values $\theta_\text{fid} - \Delta\theta_\alpha$, for each $\alpha\in\{1,\dots,p\}$. As illustrated in Fig. \ref{fig:training}, we passed the training set of images through the network to obtain the summary statistics, and we computed the quantities of interest: the covariance matrix from the fiducial set of images, the numerical derivative of the mean of the summary statistics vector for each parameter from the perturbed parameter sets, and with these, the determinant of the Fisher information matrix. For the simulations for the numerical derivatives, each parameter was varied independently, keeping the other variable parameters fixed at their fiducial values. Furthermore, for the second and third collections of images, the initial seeds were fixed to be the same for each individual image, such that the galaxies were generated at the same location in the image and with identical flux and shape. This ensured that the only changes from one image of the second collection to the corresponding image of the third collection were galaxies that appeared  or disappeared according to the luminosity function. The suppression of this variance allowed us to make fewer simulations (i.e., $m<n$) for the derivatives than for the fiducial simulations used to calculate the covariance matrix.\\
\\
At each iteration of the training phase, the training set of images was passed through the network before the back-propagation was run, that is, before the gradient of the loss function was calculated. We used the chain rule to calculate the effect of each network parameter on the value of the loss function such that they could be updated, thereby minimizing the loss. Consequently, as the training advanced, the network summarized the data while retaining increasingly more information about the model parameters. Furthermore, the network became increasingly insensitive to parameter-independent noise, providing a function that calculated extremely robust summary statistics. We used another independent set of different images that the network did not train on in order to validate the performance: the validation set.\\
\\
To update the network parameters at each iteration, we used the Adam optimizer \citep{2014arXiv1412.6980K} and/or the stochastic gradient descent method. In this training process, we did not know the expected Fisher information of the images. We therefore recommend to continue the training until the loss reaches a plateau (or starts to decrease) when evaluated on the validation set, indicating that the network has compressed the data with the maximum information available. At this point, the summaries are Gaussianly distributed in regions close to the fiducial choice of parameter values, even if the likelihood of the actual data is non-Gaussian \citep{2018PhRvD..97h3004C}.

\subsection{Choice of fiducial values: Iterative method}

If the values of the inferred parameters are far (and/or completely unknown) from the fiducial choice of training parameters, then there is a risk that the summaries are not Gaussianly distributed in the regions close to the inferred values. This can lead to very wide or flat posterior distributions when the Bayesian likelihood-free inference techniques of Appendix \ref{sec:ABC_PMC} are applied. To fix this problem, the following iterative method can be applied to choose a good set of fiducial training values:
\begin{enumerate}
    \item Train the IMNN with any guess (or random choice) for the fiducial values.
    \item Compute the quasi-maximum likelihood estimate of the data for which the parameter values are to be inferred, using Eq. \eqref{eq:MLE_summaries}.
    \item Use this estimate as the new fiducial values and train a new IMNN with these values.
\end{enumerate}
This iterative process was applied until the quasi-maximum likelihood estimate was relatively constant between iterations. It generally took very few iterations for convergence. By iterating toward the quasi-maximum likelihood, more information about the model parameters can potentially be extracted. In our applications of Sects. \ref{sec:app1} and \ref{sec:app2}, we did not need to apply this iterative method because our choice of fiducial values for the training of the network was relatively similar (or identical) to the inferred parameter values.

\section{Application to $M^*$ and $\phi^*$ for spirals alone}\label{sec:app1}

This section is only meant to explain the method for nonrealistic deep-field simulations with only one spiral population and does not allow a comparison with the CFHTLS data. However, the full description of the forward model of Sect. \ref{sec:forward} applies and stars are included as well. The goal here is to prove that the network is able to retrieve the strong correlation between the two parameters $M^*$ and $\phi^*$ of a spiral population.

\subsection{Description}

We used simulated deep fields with only one spiral population of galaxies with two free parameters of the luminosity function, $M^*$ and $\phi^*$, while the slope was fixed to $\alpha=-1.3$ and there was no redshift evolution in the luminosity function, $M_\text{evol}^*=0$ and $\phi_\text{evol}^*=0$ (but the bulge or disk redshift evolutions of Sects. \ref{sec:bulge} and \ref{sec:disk} still applied). We used a $B/T=0.2$, with the ``E'' SED for the bulge and the ``Irr'' SED for the disk, see Fig. \ref{fig:SED}. As explained in Sect. \ref{sec:training}, we chose the following fiducial parameter values and their offsets for the numerical derivatives to fit the network:
\begin{equation}
  \begin{array}{lll}      
    \theta_{\text{fid},1}&=M^*&=-20\\
    \Delta\theta_1&=\Delta M^*&=0.5\\
    \theta_{\text{fid},2}&=\log_{10}\left(\phi^*\right)&=-2.01\\
    \Delta\theta_2&=\Delta\log_{10}\left(\phi^*\right)&=0.2,\\
  \end{array}
  \label{eq:fid-sp}
\end{equation}
where $\phi^*$ is given for $H_0=100$ kms$^{-1}$Mpc$^{-1}$.
Table \ref{table:overview} gives an overview of the values that were used to generate the simulations according to the description of the forward model of Sect. \ref{sec:forward}.

\begin{figure}
    \includegraphics[width=\columnwidth]{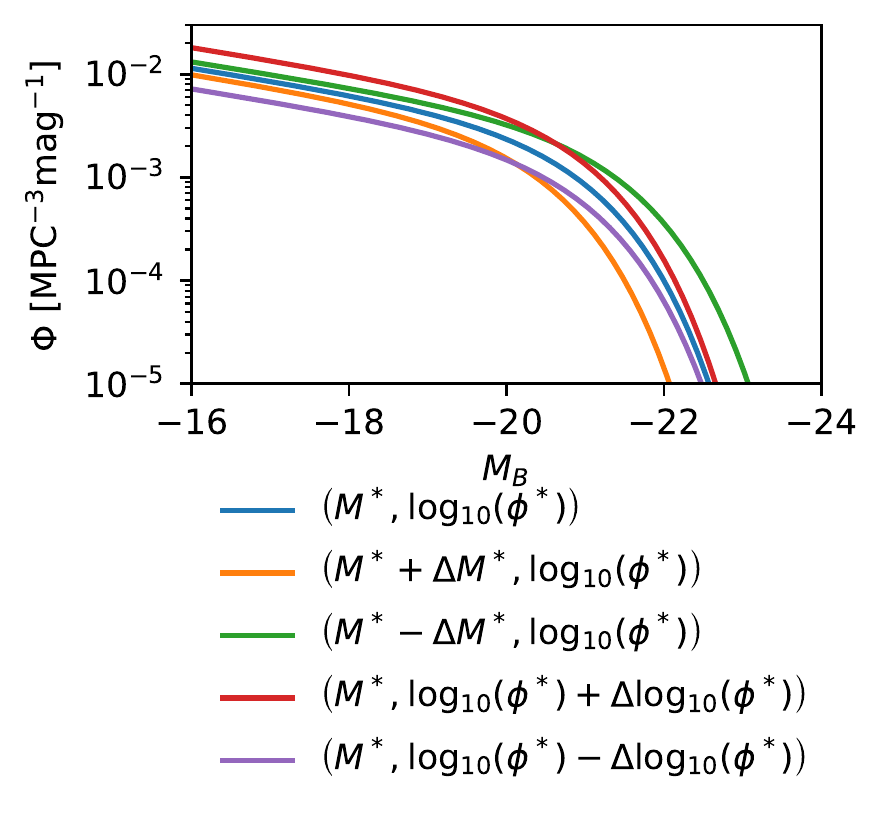}
    \caption{Theoretical luminosity functions of the spiral population with the perturbed values of the parameters used to train the network. The parameters used for each curve are listed in Eq. \eqref{eq:fid-sp}. 
    The explicit features in the data, generated from these functions, provide the differences that the IMNN will become sensitive to when training, therefore mapping the effect of $\log_{10}(\phi^*)$ and $M^*$ to informative summaries.}
    \label{fig:app1_lf}
\end{figure}

\begin{figure}
    \includegraphics[width=\columnwidth]{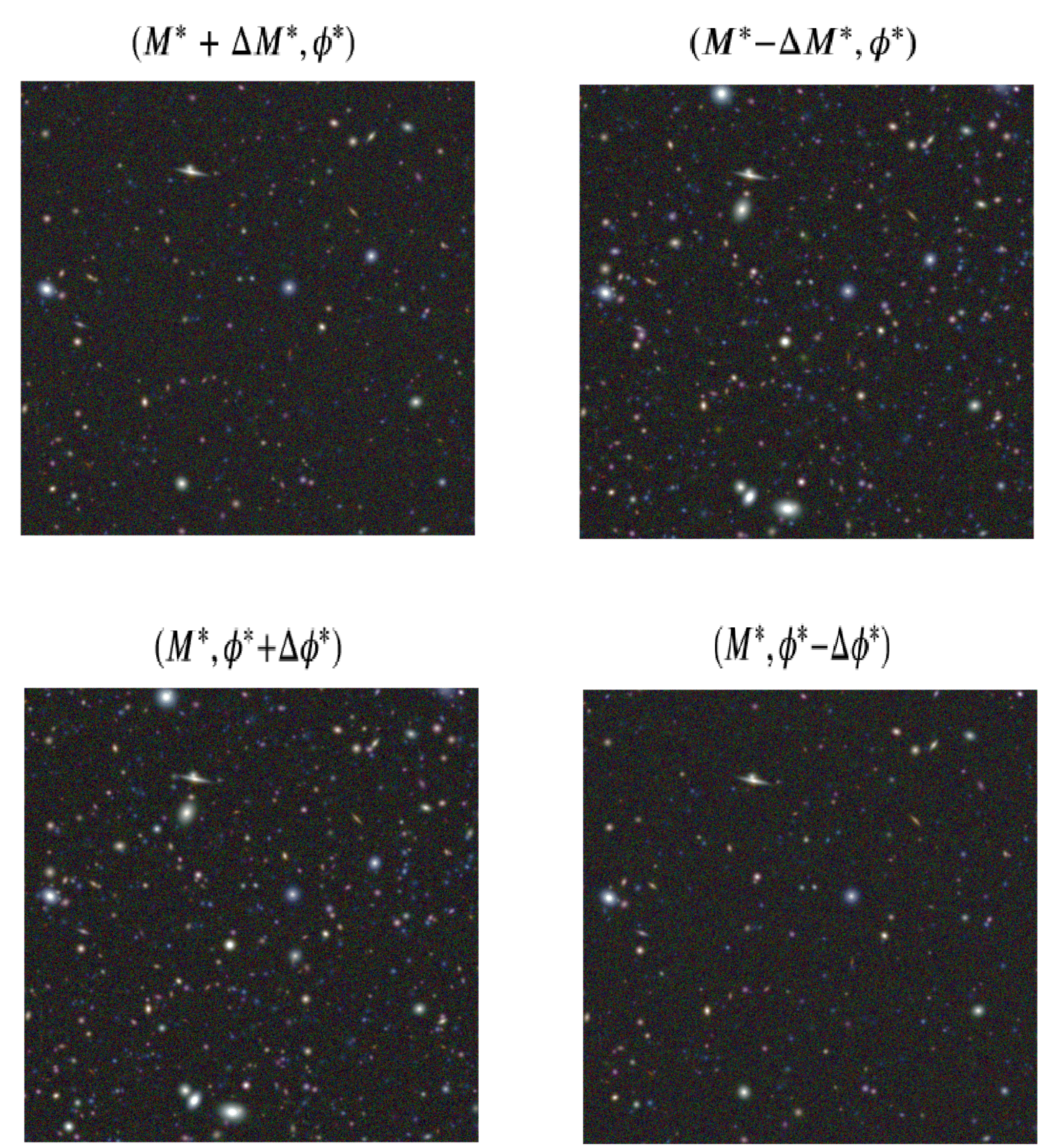}
    \caption{Color images ($g'$, $r'$, $i'$ filters) showing the effect of the perturbed values from fiducial for each parameter, as listed in Eq. \eqref{eq:fid-sp}. For each subplot we only added or removed the offset for one parameter listed in Eq. \eqref{eq:fid-sp} and kept the other at its fiducial value. Decreasing the value of $M^*$ (top right) increases the number of galaxies. Decreasing the value of $\log_{10}(\phi^*)$ decreases the number of galaxies. This shows that these two parameters are highly correlated.}
    \label{fig:app1_diff}
\end{figure}

Fig. \ref{fig:app1_lf} shows the five theoretical luminosity functions we used when making the images to train the network. Fig. \ref{fig:app1_diff} shows the effect on the RGB image (using the $i'$, $r'$, $g'$ filters) of a simulation when only one of the two parameters $M^*$ or $\phi^*$ was changed. The density of spirals is boosted for the $\Delta\log_{10}\left(\phi^*\right)$ increase in $\log_{10}\left(\phi^*\right)$ and for the $\Delta M^*$ decrease in $M^*$. Consequently, we expect a strong correlation between these two parameters.

\subsection{Training the network}\label{sec:app1_training}

To fit the network parameters (i.e., weights and biases), we followed the prescription described in Sect. \ref{sec:training} and generated $n=200$ simulations (in the eight photometric bands $u'$, $g'$, $r'$, $i'$, $z'$, $J$, $H$, $Ks$) with parameters at fiducial values. For the images used for the numerical derivatives, we generated $m=50$ simulations (in each of the eight bands) for each parameter, that is, $M^*$ and $\log_{10}(\phi^*)$, at values $\theta_{\textrm{fid},\alpha}+\Delta\theta_\alpha$, $\alpha=1,2$, and another $m=50$ simulations per parameter at values $\theta_{\textrm{fid},\alpha}-\Delta\theta_\alpha$, $\alpha=1,2$, see Eq. \eqref{eq:fid-sp}. This yielded a total of $200+2\times2\times50=400$ images (in the eight bands) for the training set, and we used the same number for the validation set. Each simulation was a $1024\times 1024$ pixel deep-field image in the $\text{eight}$ photometric bands. The optimizer techniques and the corresponding learning rates used during the training are listed in Table \ref{table:app1_lr}. We used this training scheme to quickly find a transformation that forces the covariance matrix of the summaries to be approximately the identity matrix, and then explore the more complex landscape of the parameter-dependent information. The architecture of the network is that of Fig. \ref{fig:architecture}, which yields a total of 12,170 network parameters to be fitted. Because of the number and size of the images, with this configuration, one epoch of training (i.e., when the whole training and validation sets are passed through the network and the network parameters are updated) took about 40 seconds to run (on a NVIDIA QUADRO RTX 8000 45GB GPU).

\begin{figure}
    \includegraphics[width=\columnwidth]{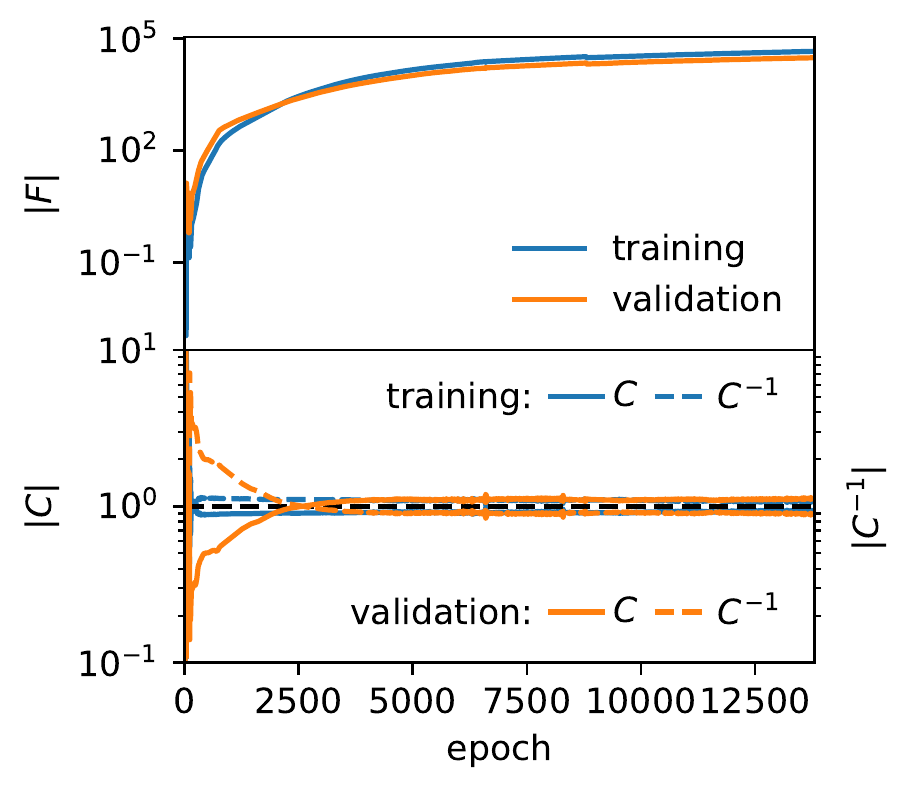}
    \caption{\textbf{Top:} Evolution of the determinant of the Fisher information matrix during almost 14000 epochs. The blue curve represents the information extracted from the training set, and the orange curve the information from the validation set. The determinant of the Fisher information is maximized on the training set and is comparable to the determinant of the Fisher information calculated with the validation set. This confirms that the two sets are representative samples. The training is stopped when the Fisher information of the validation set becomes relatively constant.\newline
    \textbf{Bottom:} Evolution of the determinants of the covariance matrix (solid line) and of the inverse of the covariance matrix (dashed line) for the training set (blue) and for the validation set (orange). The values quickly become constant, which shows that the network suppresses the parameter dependence of the covariance. Because the correlation between the parameters is strong, the covariance matrix cannot exactly diagonalize to the identity matrix, but the fact that it is constant shows that the parameter dependence is weak, which is the only requirement for the IMNN.}
    \label{fig:app1_train}
\end{figure}

We trained the network for almost 14000 epochs ($\approx$ 6 days). Fig. \ref{fig:app1_train} shows the evolution of the determinant of the Fisher information matrix (top) and the determinants of the covariance and inverse covariance matrices (bottom). The determinant of the Fisher information matrix increased during the training, and the determinant of the covariance matrix and its inverse became constant after a few hundred epochs. This shows that the covariance has very little parameter dependence, and the sensitivity in the network instead is due to the derivative of the mean of the summaries with respect to the parameters. The determinant of the covariance matrix (and its inverse) are not precisely the identity, as the regularization attempts to enforce in the optimization routine, because the parameters are extremely highly correlated, and so the summaries are as well. However, the only requirement for the IMNN is that the covariance matrix is parameter independent, and only the arbitrary scale of the summaries is set by the value of the covariance, which can be seen by the constant value in Fig.~ \ref{fig:app1_train}. The determinant of the Fisher information begins to plateau (in the training and validation data), and we decided that this is sufficient to stop the training and accept this as the amount of information we are willing to extract from the data given the training time.

\subsection{ABC posteriors}

\begin{figure}
    \includegraphics[width=\columnwidth]{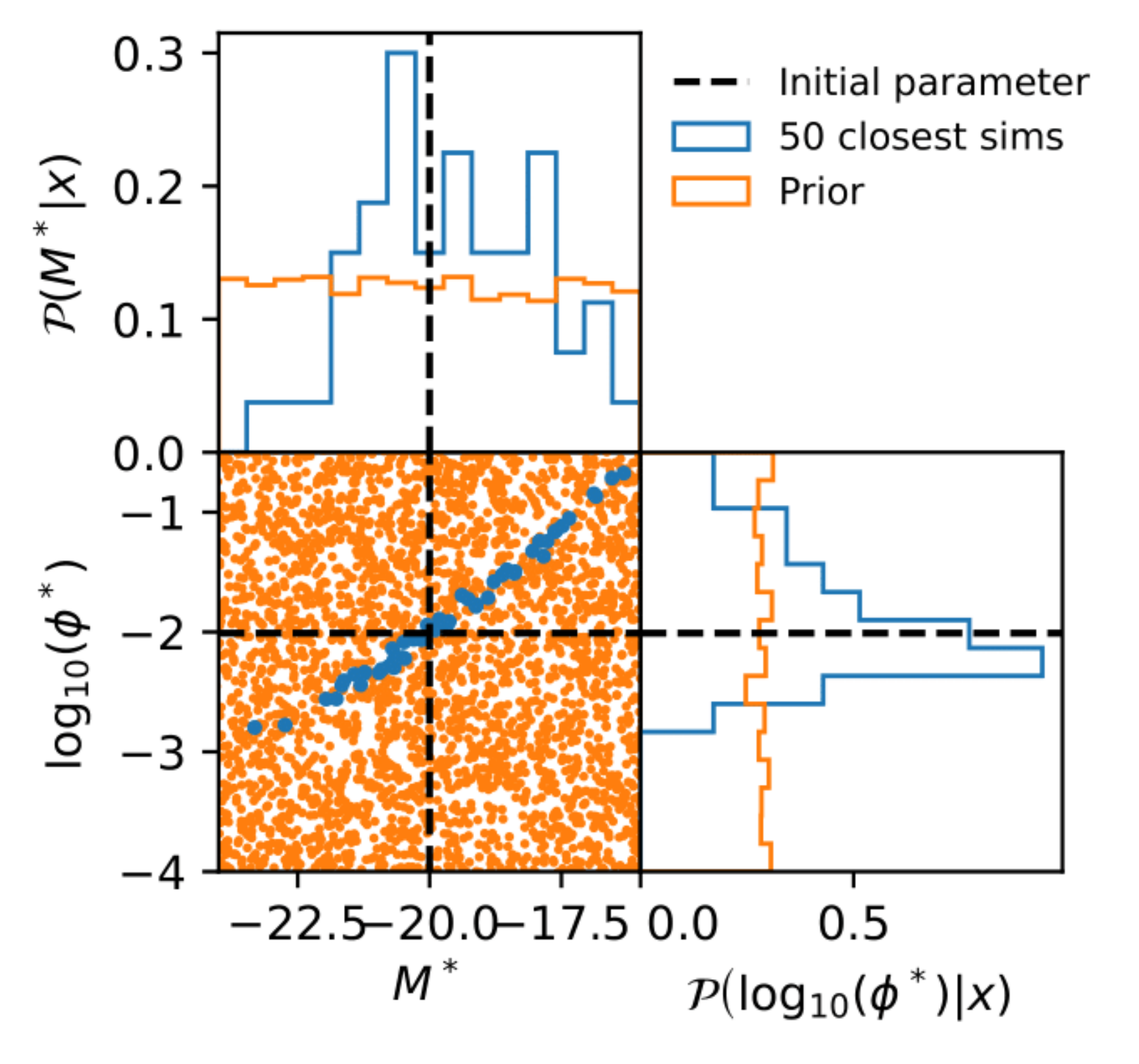}
    \caption{{\bf Results of the ABC procedure.}\newline
    {\bf Bottom left:} Values of the parameters for the 5000 prior simulations (orange) and for the $50$ simulations with minimum distance (blue) to the virtual data. The dotted red lines are the parameter values at which the virtual data were generated. We retrieve the strong correlation between the two parameters.\newline
    {\bf Top left and bottom right:} 1D marginal distributions of the parameter values.}
    \label{fig:app1_ABC}
\end{figure}

After the network was trained, we ran an ABC procedure to test its reliability, see Appendix \ref{sec:ABC} for more details. Here, we replaced what is referred to as "observed data" in Appendix \ref{sec:ABC} with a simulation called "virtual data", generated with parameters set to the fiducial parameter values in Eq. \eqref{eq:fid-sp},  
\begin{equation}
    \begin{array}{lll}   
    \theta_{\text{vir},1}&=M^*&=-20\\
    \theta_{\text{vir},2}&=\log_{10}\left(\phi^*\right)&=-2.01,
    \end{array}
    \label{eq:sim-sp}
\end{equation}
where $\phi^*$ is given for $H_0=100$ kms$^{-1}$Mpc$^{-1}$. We used the formalism of Appendix \ref{sec:ABC} and chose a uniform prior for the two parameters,
\begin{equation}
    \begin{array}{rl}   
    M^*&\in[-24,-16]\\
    \log_{10}\left(\phi^*\right)&\in[-4,-0.5].\\
    \end{array}
    \label{eq:sim-Sp-prior}
\end{equation}

From these prior intervals, we uniformly drew an initial sample of 5000 pairs of parameters and generated a simulated deep field (using the forward model of Sect. \ref{sec:forward}) for each pair. We then passed both the virtual data and the 5000 simulated fields through the network in order to compress them. The bottom left plot of Fig. \ref{fig:app1_ABC} shows the values of these parameters for the entire 5000 prior simulations in orange. We decided to only keep the $50$ simulations in blue for which the distance to the virtual data is minimal. The dotted red lines are the parameter values used to generate the virtual data. The ABC procedure retrieves the expected strong correlation between the two parameters. However, because we do not know which value of $\epsilon$ should be used, the posterior is not properly approximated here, therefore we used the population Monte Carlo (PMC) procedure of Appendix \ref{sec:PMC} to do it automatically. However, the ABC procedure works as well and gives a clear indication of why the method works.

\subsection{PMC posteriors}

\begin{figure}
    \includegraphics[width=\columnwidth]{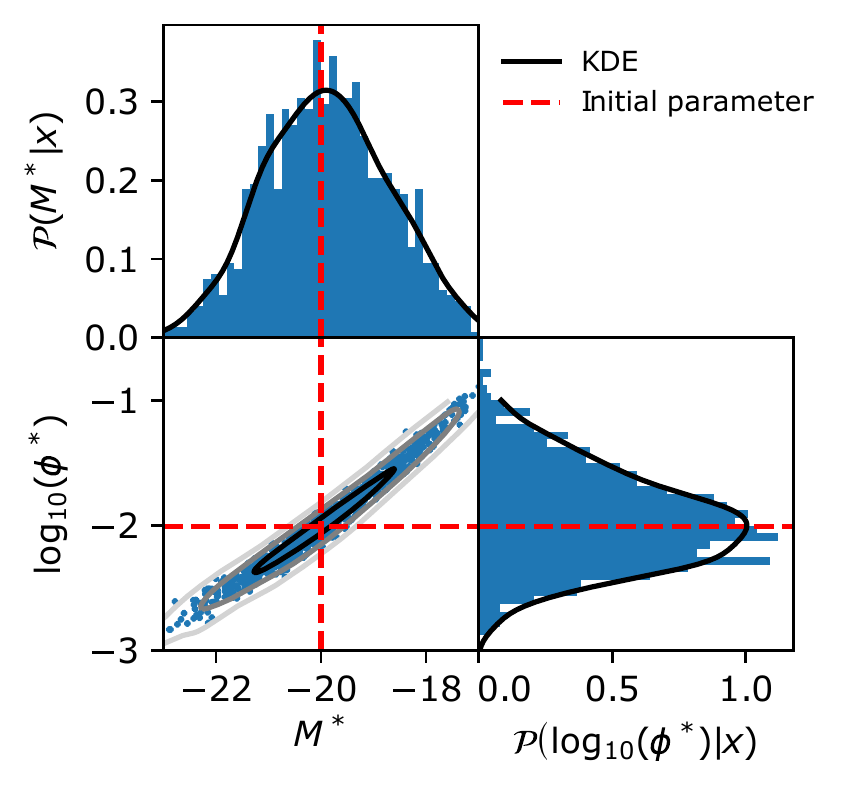}
    \caption{{\bf Results of the PMC procedure.}\newline
    {\bf Bottom left:} Distribution of the parameter values for final 1000 points obtained by the PMC. The black and dark, gray and light, and  gray show the $68\%,95\%,\text{ and }99\%$ contours, and the dotted red line shows the values of the parameters that were used to generate the virtual data.\newline
    {\bf Top left and bottom right:} 1D marginal distributions of the parameters and their KDE in black.\newline
    The procedure converged around the parameter values we used to generate the virtual data. There is evidence here, in 2D, that the uncertainty due to the correlation between the parameters whose effect is evident in the data is large, see Fig. \ref{fig:app1_diff}.}
    \label{fig:app1_PMC}
\end{figure}

For the PMC, we used the same prior for the two parameters as in Eq. \eqref{eq:sim-Sp-prior}, and we applied the PMC procedure of Appendix \ref{sec:PMC} with $N=1000,$  $M=50000,$ and $Q=75\%$ for resampling the $N$ sets of parameters. This yielded about 232577 draws and generated simulations in total. The results are shown in Fig. \ref{fig:app1_PMC}. The 2D distribution (bottom left) and the 1D projected distributions are concentrated around the values of the parameters we used to generate the virtual data (i.e., $M^*=-20$ and $\log_{10}(\phi^*)=-2.01$). The procedure clearly identifies the strong correlation between the two parameters, both affecting the density of objects, and is able to infer that the parameter values we used to generate the virtual data lie in the bulk of the probability density. This application illustrates that the network is able to summarize the effects in the images of the two correlated parameters, and that it is able to robustly constrain the possible values of the parameters we used to generate the virtual data.

\section{Application to $\phi_0^*$ for ellipticals and spirals}\label{sec:app2}

In this section, we consider a more realistic model with two populations of galaxies, ellipticals and spirals, which can then be compared to the observed data for the CFHTLS and WIRDS D1 deep field. We only infer the density parameter $\phi_0^*$ of the luminosity functions for these two populations, and we use the recent results of \citet{2017A&A...599A..62L} as a reference because they provided a deep statistical analysis of the luminosity function parameters for these two populations of galaxies.

\subsection{Description}

\begin{figure}
    \includegraphics[width=\columnwidth]{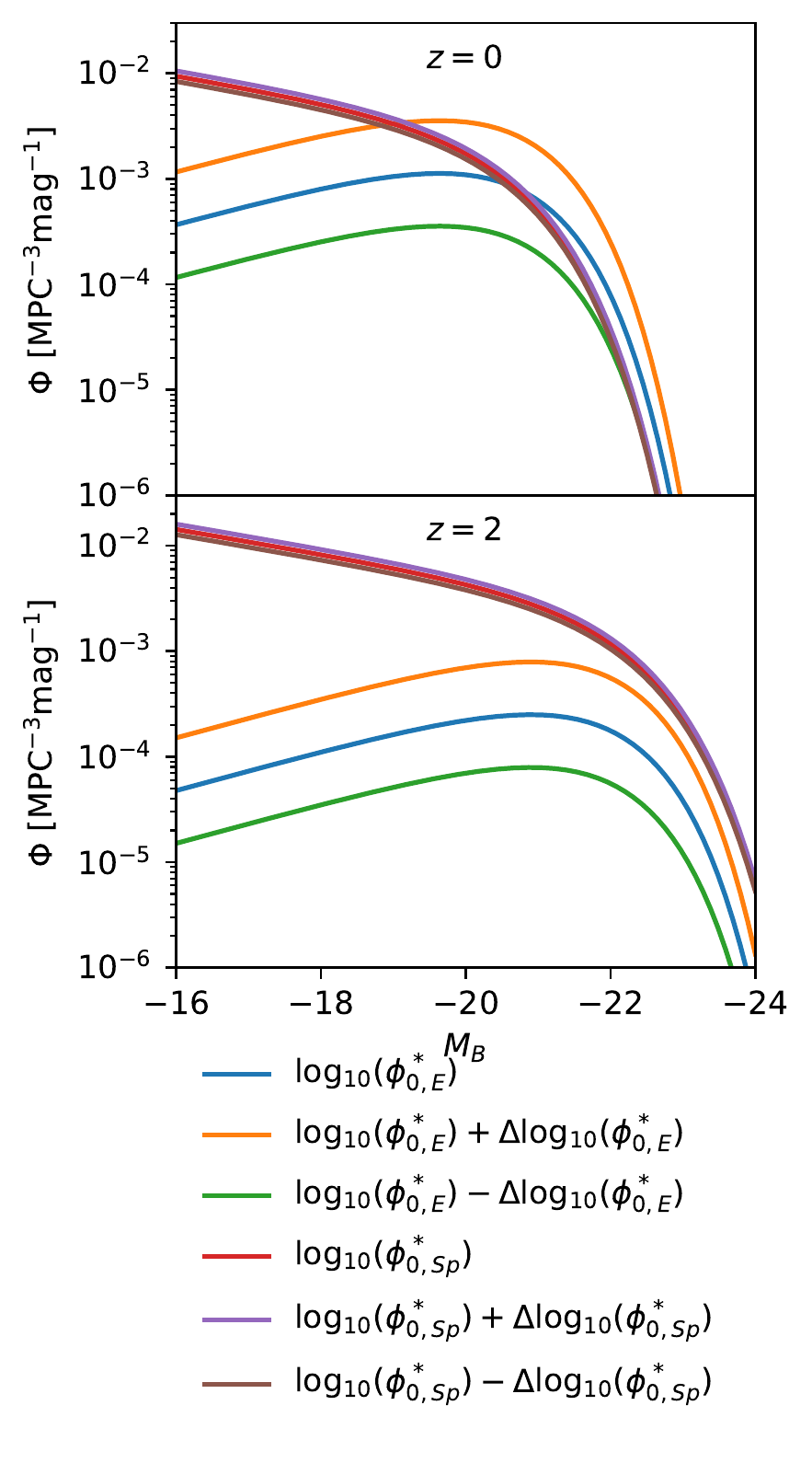}
    \caption{Theoretical luminosity functions of the elliptical (blue, orange, and green) and spiral (red, purple, and brown) populations at redshift $z=0$ (top) and $z=2$ (bottom). The legend applies to both panels. The parameters used for each luminosity function are given in Table \ref{table:overview}.
    These curves show the higher integrated density of spiral galaxies than of elliptical galaxies. The steep faint-end slope of the spiral population implies that the images contain many faint spiral galaxies, which is not the case for the elliptical population. There is also a redshift effect that decreases the proportion of ellipticals or spirals when looking at distant objects.}
    \label{fig:app2_lf_evol}
\end{figure}

\begin{figure*}
\centering
\includegraphics[scale=0.65]{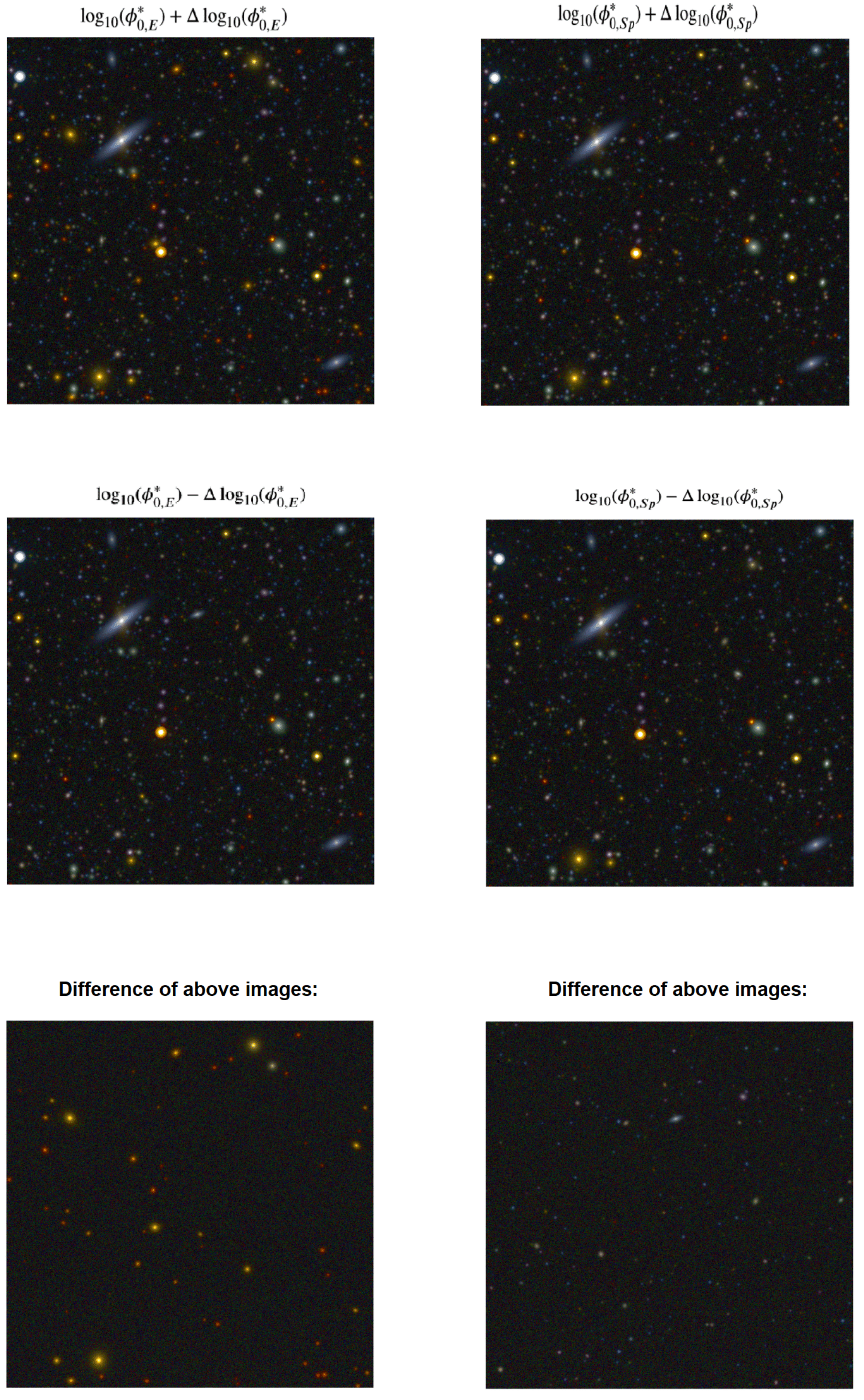}
\caption{{\bf Left:} Difference of the two upper images is shown in the bottom panel.\ The two upper images are simulated by only changing the fiducial parameter of the elliptical population to the perturbed parameter values by $+\Delta\theta_1$ (top) and $-\Delta\theta_1$ (middle). Only yellow or red elliptical galaxies remain, which confirms that $\phi^*_{0,E}$ affects the density of ellipticals in the simulations.\newline
{\bf Right:} Difference of the two upper images is shown in the bottom panel. The two upper images are simulated by only changing the fiducial parameter of the spiral population to the perturbed parameter values by $+\Delta\theta_2$ (top) and $-\Delta\theta_2$ (middle). Only blue spiral galaxies remain, which confirms that $\phi^*_{0,Sp}$ affects the density of spirals in the simulations. These RGB color images are obtained from the $i'$, $r'$, $g'$ filters.}
\label{fig:diff_params}
\end{figure*}

We used the following fiducial parameter values from \citet{2017A&A...599A..62L} and offsets for the numerical derivatives to fit the network:
\begin{equation}
  \begin{array}{lll}      
    \theta_{\text{fid},1}&=\log_{10}\left(\phi_{0,E}^*\right)&=-2.09\\
    \Delta\theta_1&=\Delta \log_{10}\left(\phi_{0,E}^*\right)&=0.5\\
    \theta_{\text{fid},2}&=\log_{10}\left(\phi_{0,Sp}^*\right)&=-2.04\\
    \Delta\theta_2&=\Delta\log_{10}\left(\phi_{0,Sp}^*\right)&=0.05,\\
  \end{array}
  \label{eq:fid-e-sp}
\end{equation}
where $\phi^*$ is given for $H_0=100$ kms$^{-1}$Mpc$^{-1}$. The other luminosity function parameters were set to be constant with the values from \citet{2017A&A...599A..62L}, see Table \ref{table:overview}. We considered an offset value of $\phi_{0,\textrm{Sp}}^*$ a factor of 10 smaller for the spiral galaxies than $\phi_{0,\textrm{E}}^*$ for the ellipticals when we generated the simulations for the numerical derivatives, see Eq. \eqref{eq:fid-e-sp} because the density of spirals is higher than that of ellipticals, see Fig. \ref{fig:app2_lf_evol}. We used the reference SDSS $g'$ band for the absolute magnitude, therefore we applied a correction of $B-g'=0.78$ for the elliptical population and $B-g'=0.40$ for the spiral population, as suggested by Table 7 of \citet{1995PASP..107..945F}, to express the magnitudes in the reference $B$ band.\\
\\
For computational efficiency, we only used a 3.17 arcmin$^2$ images (corresponding to $1024\times1024$ pixels of 0.186 arcsec) of the full 1 deg$^2$ D1 deep field in the eight photometric bands $u'$, $g'$, $r'$, $i'$, $z'$, $J$, $H$, $Ks$ to infer the values of $\phi_0^*$ for the elliptical and spiral populations. Fig. \ref{fig:diff_params} shows the effects on a color image of the $g'$, $r'$, $i'$ simulations for the elliptical population (left column) and the spiral population (right column) of changing the fiducial parameters by their respective positive (top images) and negative (middle images) offset from the fiducials that were used to calculate the numerical derivatives. The bottom image of each column is the difference of the above images and shows the galaxies that appear in the images we used to calculate the numerical derivatives of the summaries using the IMMN. The bottom left panel shows that only elliptical galaxies remain when only $\phi_0^*$ for the elliptical is changed (these galaxies appear as yellow or red because of the SED used to model them), as opposed to the blue galaxies of the spiral population remaining in the bottom right panel. Table \ref{table:overview} gives an overview of the values that were used to generate the simulations according to the description of the forward model of Sect. \ref{sec:forward}.

Fig. \ref{fig:app2_lf_evol} shows the six theoretical luminosity functions, at $z=0$ and $z=2$, that we used to train the network. The density of bright red ellipticals is lower at $z=2$ than at $z=0$: fewer galaxies with early-type SEDs existed in the past than today because they were still forming their stars and therefore had bluer SEDs. We also expect a relatively higher density of faint spirals compared to ellipticals at $z=0$ and $z=2$. The curves for the perturbed luminosity functions of spirals are closer together because the offset $\Delta\theta$ is a factor 10 lower than for the ellipticals.

\subsection{Training of the network}
\label{sec:app2_training}

\begin{figure}
    \includegraphics[width=\columnwidth]{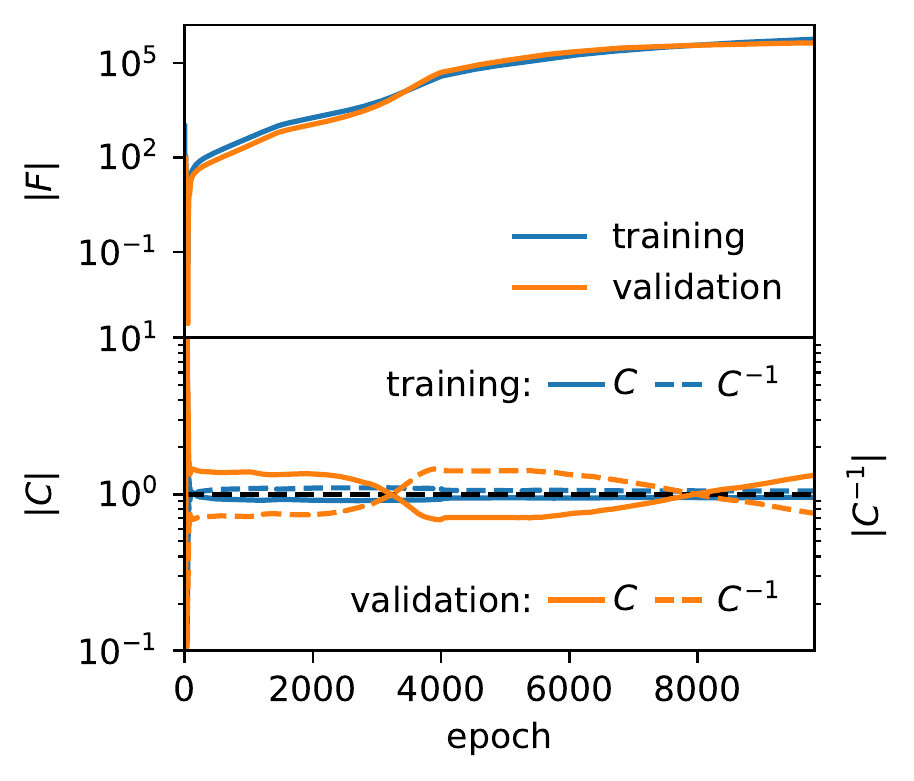}
    \caption{\textbf{Top:} Evolution of the determinant of the Fisher information matrix during almost 10 000 epochs. The blue curve represents the information extracted from the training set, and the orange curve shows the same from the validation set. The curves increase, which means that the network learns more about the two parameters as the training continues. The training was stopped when the validation curve flattened, suggesting that the network has converged.\newline
    \textbf{Bottom:} Evolution of the determinant of the covariance matrix (solid line) and the inverse of the covariance matrix (dashed line). The blue curves show the training set, and the orange curves show the validation set. The training curves reach 1 very fast, which shows that the loss is stabilized and that the magnitude of the summaries is under control. The validation curve oscillates while being still very close to identity, which is a sign that there is some weak parameter dependence on the covariance.}
    \label{fig:app2_train}
\end{figure}

\begin{figure*}
\centering
\includegraphics[width=\textwidth]{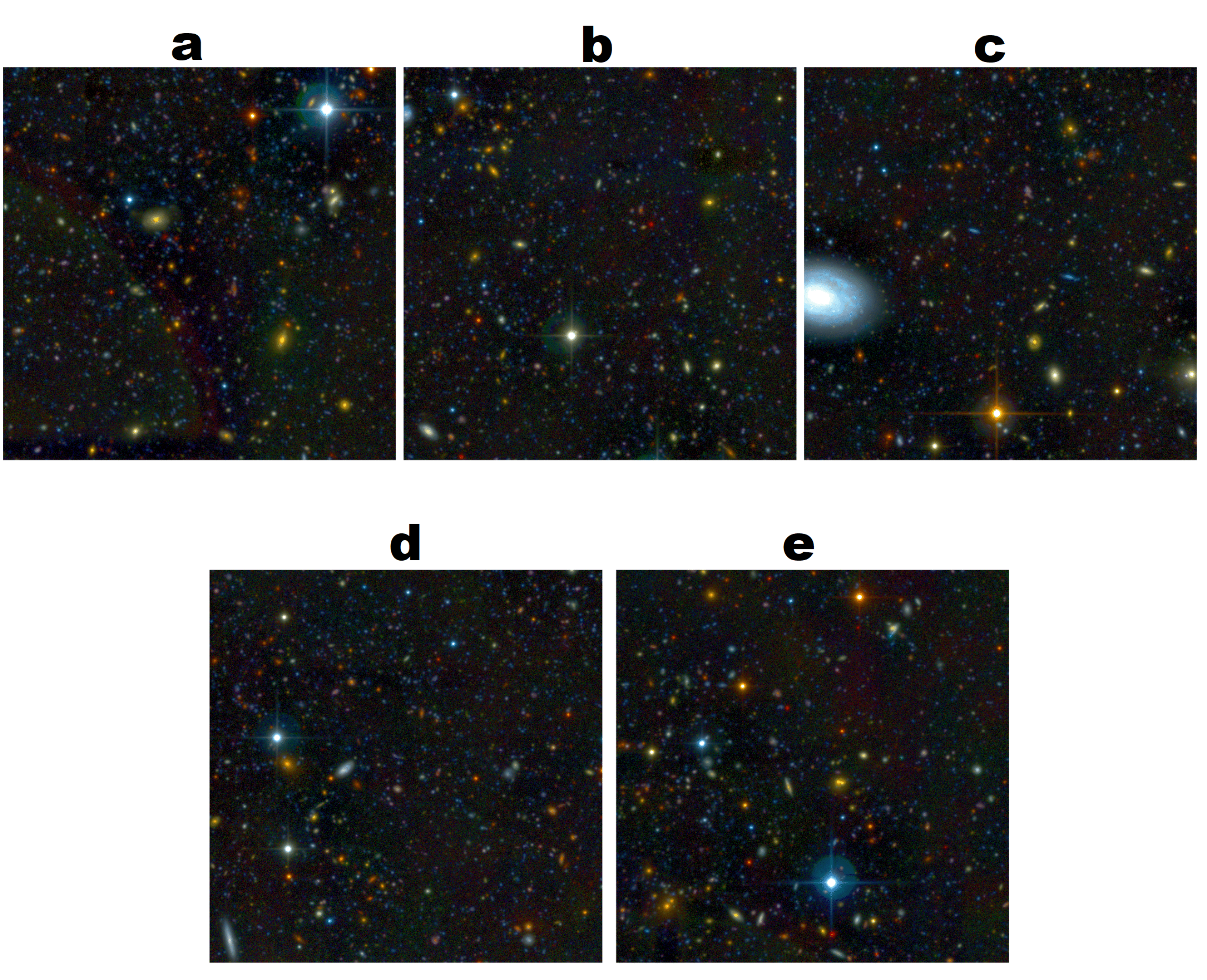}
\caption{RGB images using the $g'$, $r'$, $i'$ filters for five random regions of 3.17 arcmin$^2$ within the 1 deg$^2$ CFHTLS D1 deep field that are used to infer the luminosity function parameters of the elliptical and spiral galaxies, namely the logarithm of their amplitudes $\log_{10}(\phi_{0,\textrm{E}}^*)$ and $\log_{10}(\phi_{0,\textrm{Sp}}^*)$.}
\label{fig:app2_D1}
\end{figure*}

\begin{figure*}
\centering
\includegraphics[width=\textwidth]{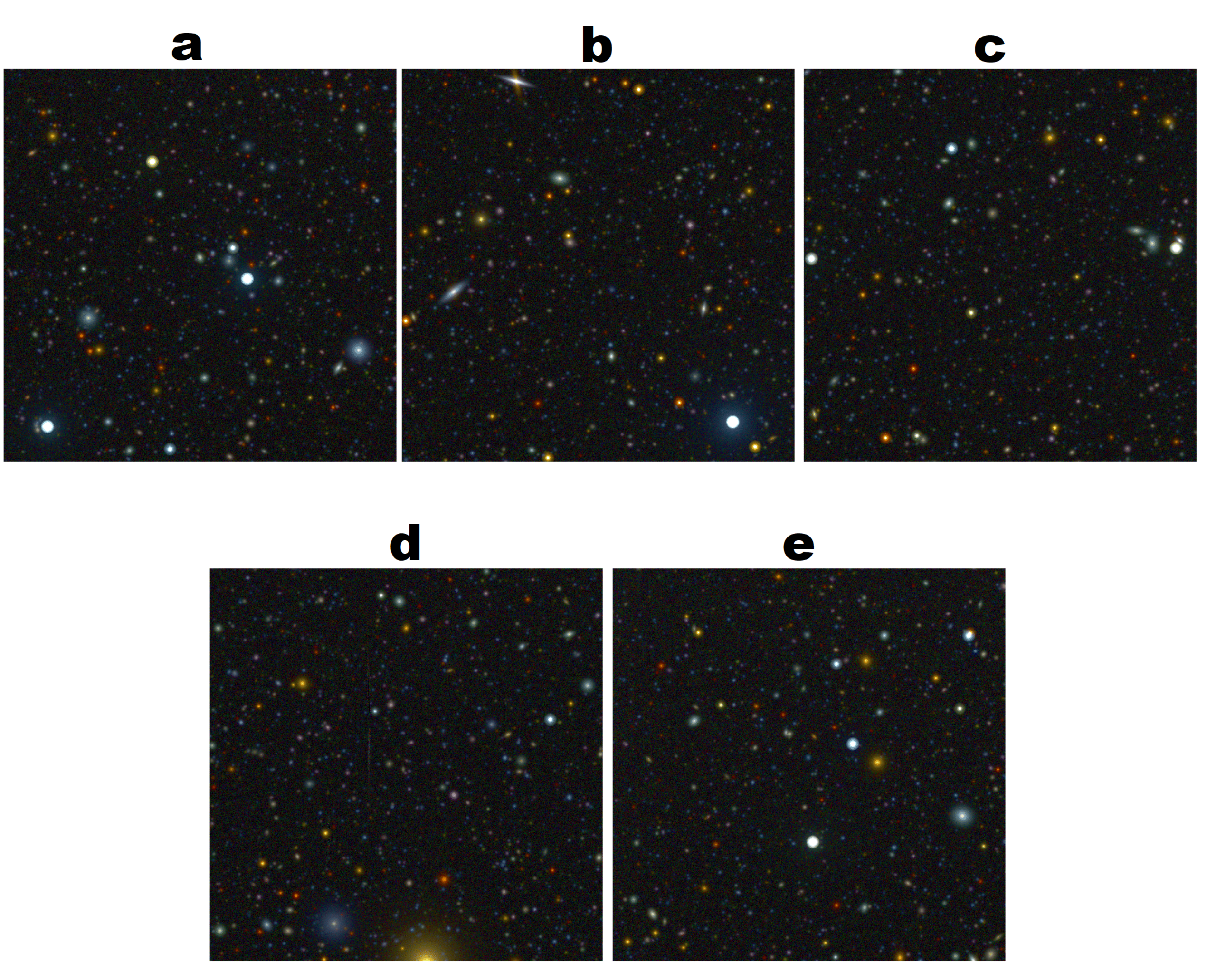}
\caption{RGB images using the $g'$, $r'$, $i'$ filters for the five virtual data of 3.17 arcmin$^2$ generated with fiducial values of the luminosity function parameters (see Table \ref{table:overview}), used to validate the method.}
\label{fig:app2_fake}
\end{figure*}

As in the previous application of Sect. \ref{sec:app1_training}, we generated simulations in the same way, with $n=200$ fiducial images of 3.17 arcmin$^2$ and $1024\times1024$ pixels in the $\text{eight}$ photometric bands, and $2\times2\times50=200$ ($m=50$) seed-matched images for the numerical derivatives. We used the same inception architecture with the Adam optimizer (learning rate of 0.04) for the first 4000 epochs of the training, then the stochastic gradient descent optimizer (learning rate of $5\times10^{-6}$) for the remaining epochs. We trained the network on the GPU NVIDIA TITAN X 12Go for almost 10 000 epochs ($\approx$ 8 days), and we show in Fig. \ref{fig:app2_train} the convergence of the determinant of the Fisher matrix (top) and the determinant of the covariance and inverse covariance matrices (bottom) for the training set (blue) and the validation set (orange). During the fitting, the training and validation curves remain close to each other, which indicates that the training and validation sets contain approximately the same amount of information about the two parameters. Consequently, both sets seem representative enough of the diversity of realizations of a deep field of spiral and elliptical galaxies with these fiducial parameters.

\subsection{Observed and virtual data}

Our ultimate goal is to infer the parameters on the 1 deg$^2$ D1 deep field. However, we limited the analysis to deep fields of size 3.17 arcmin$^2$ (i.e., $1024\times 1024$ pixels) for computational efficiency, and randomly chose five such regions of the D1, see Fig. \ref{fig:app2_D1}. In order to confirm that the network is trained properly and performs well, we also generated five simulations with the fiducial parameters of Table \ref{table:overview} and the same angular size as the CFHTLS data, which we consider as virtual data (see Fig. \ref{fig:app2_fake}). The goal is to obtain the individual posterior distributions of the five observed data images, then to compute a joint posterior to constrain the values of the parameters and extract the confidence intervals. The same procedure was applied to the five virtual data. The difference was that we knew the values of the parameters that were used to generate the virtual data. We chose five images in both cases as a good compromise between the computational time and the accuracy of the joint posterior.

\subsection{ABC posteriors}

\begin{figure}
    \includegraphics[width=0.5\textwidth]{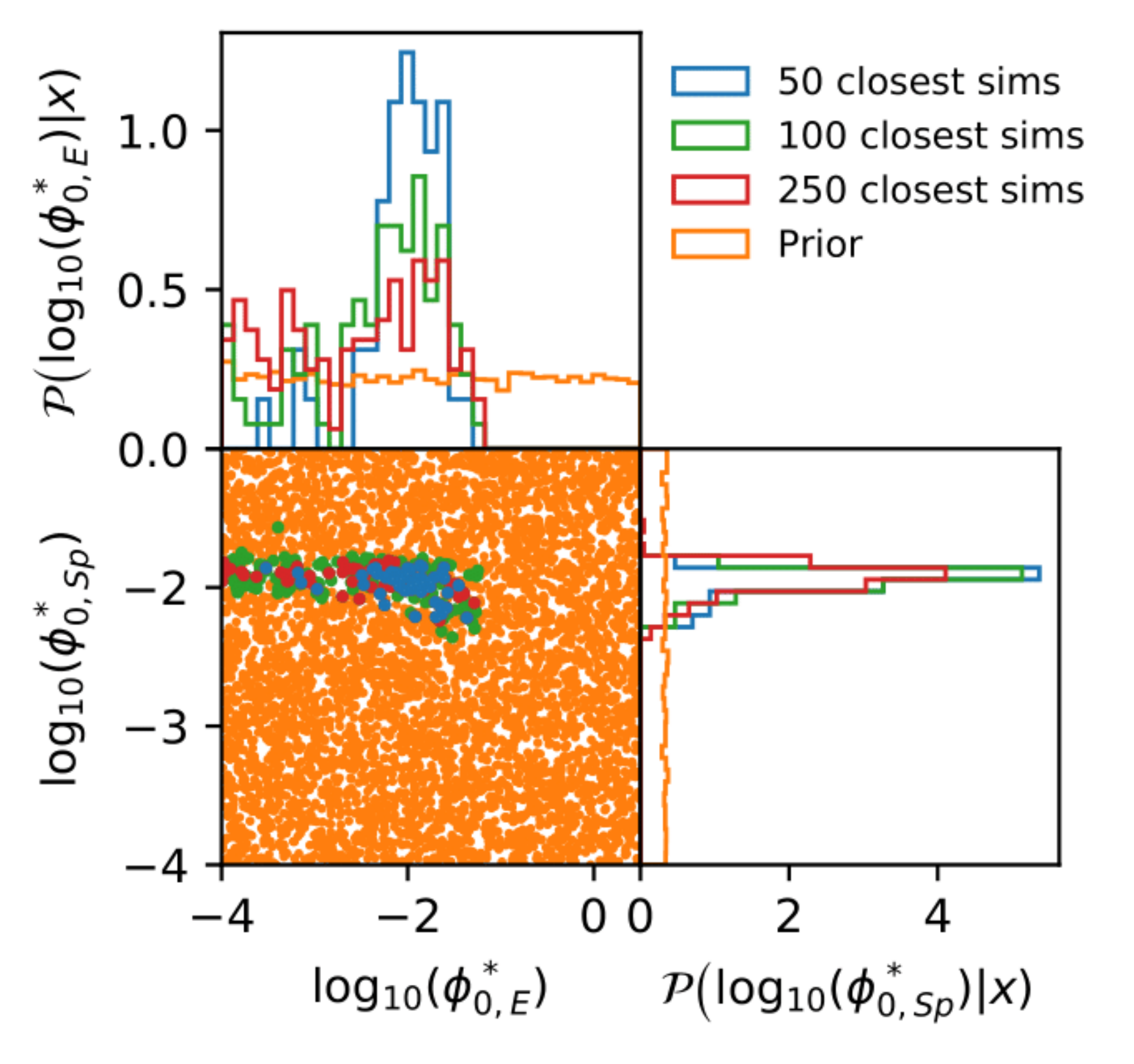}
    \caption{{\bf Results of the ABC procedure.}\newline
    {\bf Bottom left:} Parameter values corresponding to 5000 simulations (dots) drawn from our random uniform prior (orange) of Eq. \eqref{eq:sim-E-Sp-prior}.
    The colored points are those with a small distance $\rho$ (Eq. \eqref{eq:dist}) to the observed data (frame 'b' of Fig. \ref{fig:app2_D1} from the CFHTLS D1 Deep field): the $50$ closest points are shown in blue, the $100$ closest points are shown in green, and the $250$ closest points are shown in red. \newline
    {\bf Top left and bottom right:} Marginalized distributions of the distance selections in the bottom left panel. The points with the smallest distances appear to be around $\left(\log_{10}(\phi_{0,\textrm{E}}^*),\log_{10}(\phi_{0,\textrm{Sp}}^*)\right)=(-2,-2)$.}
    \label{fig:app2_ABC}
\end{figure}

We applied the ABC procedure of Appendix \ref{sec:ABC} and used a uniform prior distribution with the same lower bound for the amplitudes of the luminosity functions of the elliptical and spirals galaxies, but different upper bounds:
\begin{equation}
    \begin{array}{rl}   
    \log_{10}(\phi^*_{0,\textrm{E}})&\in[-4,0.5]\\
    \log_{10}(\phi^*_{0,\textrm{Sp}})&\in[-4,-1].\\
    \end{array}
    \label{eq:sim-E-Sp-prior}
\end{equation}
The choice of a smaller upper bound of the prior for the spirals was made because the luminosity function for this population has a steep faint end slope (see Table \ref{table:overview} and Fig. \ref{fig:app2_lf_evol}), and it can lead to very long generation times of the simulations with high values of the amplitude $\phi^*_{0,\textrm{Sp}}$. Moreover, previous studies such as \citet{2017A&A...599A..62L} have shown that $\log_{10}(\phi_{0,\textrm{Sp}}^*)>-1$ is very unlikely, and indeed, this is what we also find from Fig. \ref{fig:app2_PMC_real}. We wish to note that cutting possible regions of parameter space due to computational resources is not a statistically rigorous process, but the information from  \citet{2017A&A...599A..62L} gives us reason to believe that it is acceptable to truncate here.

Fig. \ref{fig:app2_ABC} shows the results of the ABC procedure with 5000 draws of the parameters following the uniform prior of Eq. \eqref{eq:sim-E-Sp-prior}. A distance threshold can be chosen from the data to accept or reject some of the parameters, as shown by the green, blue, and red points in the bottom left plot. The number of accepted points decreases when this threshold is low, but their 2D region also becomes more concentrated. As a consequence, the corresponding 1D marginalized distribution of $\log_{10}(\phi_{0,E}^*)$ (top left) varies significantly depending on this choice for a distance threshold. This is because we did not densely sample from a stationary distribution, a problem that can occur when sparsely drawing from the prior and not choosing a small enough $\epsilon-$ball in which to accept simulations. In order to properly approximate the posterior, the distance threshold must approach $\epsilon=0$ and more simulations need to be done for the ABC until a steady distribution is reached. This highlights the need for a PMC that automatically approaches the posterior, which we describe in the following.

\subsection{PMC posteriors and confidence intervals}

\begin{figure}
    \includegraphics[width=0.5\textwidth]{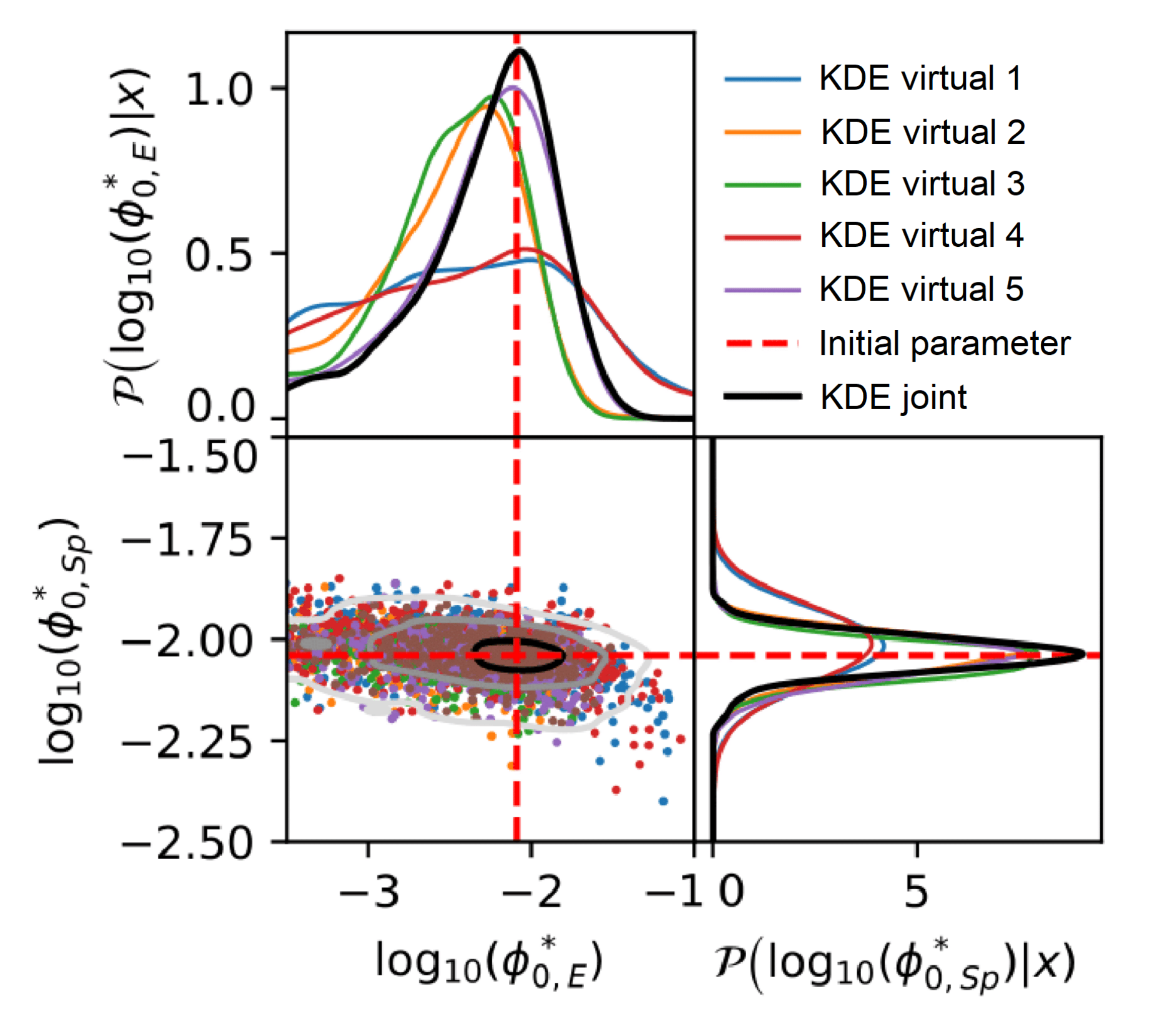}
    \caption{Posterior distributions of the two parameters $\log_{10}(\phi_{0,\textrm{E}}^*)$ and $\log_{10}(\phi_{0,\textrm{Sp}}^*)$ for the five images of virtual data (with different colors from blue to purple) and for the joint-PMC (black) described in Sect. \ref{sec:joint-pmc}. The $68\%$, $95\%,$ and $99\%$ contours of the joint-PMC are plotted in black, gray, and light gray, respectively, in the bottom left panel. The parameters used to generate the virtual data are indicated with dashed red lines. The five individual 1D posteriors for each image peak in the same region of parameter values, which indicates that the most likely parameters are consistent among the five images. The deviation between the different posteriors arises from the fact that these fields are stochastically sampled from a random process and so statistical differences exist in the data. The joint posterior is tighter and shows how likely the parameters would be if we considered the five images simultaneously.}
    \label{fig:app2_PMC_fake}
\end{figure}

\begin{figure}
    \includegraphics[width=0.5\textwidth]{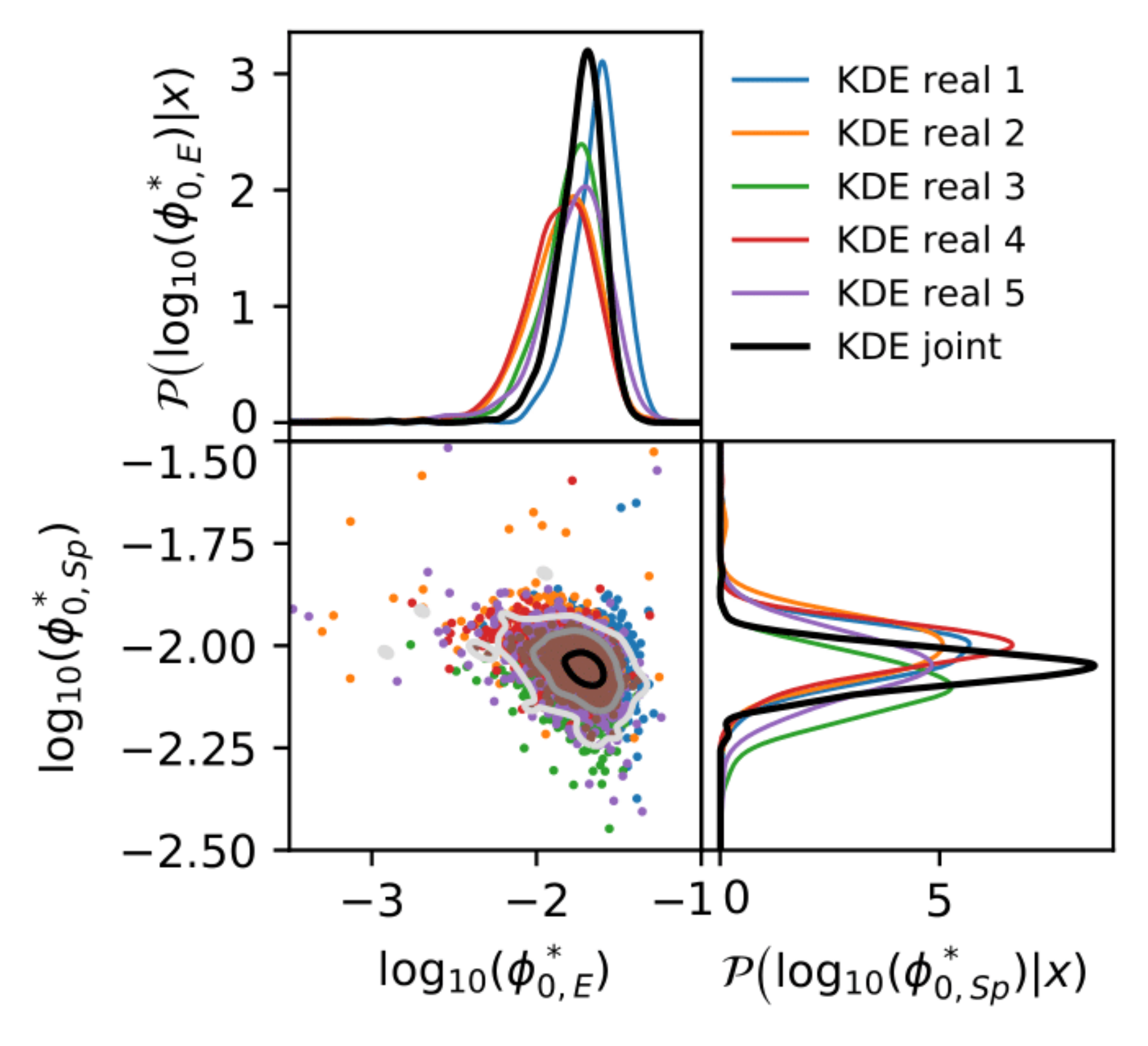}
    \caption{Posterior distributions of the two parameters $\log_{10}(\phi_{0,\textrm{E}}^*)$ and $\log_{10}(\phi_{0,\textrm{Sp}}^*)$ for the five images of observed data (blue to purple) and for the joint-PMC (black) described in Sect. \ref{sec:joint-pmc}. The $68\%$, $95\%,$ and $99\%$ contours of the joint-PMC are plotted in black, gray, and light gray, respectively, in the bottom left panel. The five individual 1D posteriors for each inset peak in the same region of parameter values, which indicates that the most likely parameters are consistent among the five images. The differences in the posteriors obtained from the different images come from the fact that the observed data come from different patches of the sky with statistically different amounts of information in the patches due to their independent environments. The joint posterior is tighter and shows how likely the parameters would be if we considered the five images simultaneously.}
    \label{fig:app2_PMC_real}
\end{figure}

We applied a parallelized version of the PMC procedure to the five images of the virtual and observed data. We used the same priors as in Eq. \eqref{eq:sim-E-Sp-prior} and applied the PMC procedure of Appendix \ref{sec:PMC} with $N=500,$  $M=5000,$ and a $25\%$ threshold for resampling the $N$ sets of parameters. This yielded about 60 000 draws and generated simulations per image. 

The results are shown in Fig. \ref{fig:app2_PMC_fake} for the virtual data and Fig. \ref{fig:app2_PMC_real} for the observed data. As shown by the 2D distribution of points (bottom left) and the 1D projected Gaussian kernel density estimates, the posteriors are concentrated around similar regions of parameter values for the five insets of each sample for the observed and virtual data because the data come from the same generative process, while the differences in the posteriors are due to the independent environment (or realization) containing more or less constraining power. We provide the $68\%$, $95\%,$ and $99\%$ confidence intervals for the 1D marginal distributions. Because the distribution is not Gaussian in 2D, its covariance (nor the 1D marginal standard deviations) are sufficient to wholly encapsulate the information in this posterior distribution. This gives us further good reason to consider such likelihood-free inference methods. The parameter values used to generate the virtual data images are shown with dotted red lines in Fig. \ref{fig:app2_PMC_fake} and can  be compared to the posterior distributions. All results, namely the parameter values used to generate the data (for the virtual data alone) and the $68\%$, $95\%,$ and $99\%$ 1D confidence intervals for $\Phi^*_{0,\textrm{E}}$ and $\Phi^*_{0,\textrm{Sp}}$ are listed in Tables \ref{table:app2_fake} and \ref{table:app2_real} for the five individual images.

\begin{table*}
\caption{1D confidence intervals for the five images of the virtual data and the initial values at which the virtual data were generated.}
\label{table:app2_fake}
\begin{center}
    \begin{tabular}{c c c c c}
        \hline\hline
         & \text{Initial} & 68\% & 95\% & 99\% \\
        \hline
        \text{virtual 1} & & & & \\
        $\log_{10}(\phi_{0,\textrm{E}}^*)$ & -2.09 & [-3.20 , -1.63] & [-3.86 , -1.23] & [-3.99 , -0.93] \\
        $\log_{10}(\phi_{0,\textrm{Sp}}^*)$ & -2.04 & [-2.11 , -1.93] & [-2.23 , -1.86] & [-2.45 , -1.86] \\
        \hline
        \text{virtual 2} & & & & \\
        $\log_{10}(\phi_{0,\textrm{E}}^*)$ & -2.09 & [-2.86 , -1.94] & [-3.67 , -1.76] & [-4.00 , -1.59] \\
        $\log_{10}(\phi_{0,\textrm{Sp}}^*)$ & -2.04 & [-2.09 , -1.98] & [-2.16 , -1.93] & [-2.31 , -1.86] \\
        \hline
        \text{virtual 3} & & & & \\
        $\log_{10}(\phi_{0,\textrm{E}}^*)$ & -2.09 & [-2.78 , -1.97] & [-3.56 , -1.76] & [-3.93 , -1.75] \\
        $\log_{10}(\phi_{0,\textrm{Sp}}^*)$ & -2.04 & [-2.10 , -2.00] & [-2.15 , -1.96] & [-2.21 , -1.91] \\
        \hline
        \text{virtual 4} & & & & \\
        $\log_{10}(\phi_{0,\textrm{E}}^*)$ & -2.09 & [-3.15 , -1.60] & [-3.94 , -1.27] & [-3.99 , -0.85] \\
        $\log_{10}(\phi_{0,\textrm{Sp}}^*)$ & -2.04 & [-2.11 , -1.92] & [-2.26 , -1.85] & [-2.92 , -1.85] \\
        \hline
        \text{virtual 5} & & & & \\
        $\log_{10}(\phi_{0,\textrm{E}}^*)$ & -2.09 & [-2.60 , -1.76] & [-3.50 , -1.58] & [-3.95 , -1.58] \\
        $\log_{10}(\phi_{0,\textrm{Sp}}^*)$ & -2.04 & [-2.09 , -1.99] & [-2.17 , -1.95] & [-2.23 , -1.88] \\
        \hline
        \text{chain-rule joint} & & & & \\
        $\log_{10}(\phi_{0,\textrm{E}}^*)$ & -2.09 & [-2.55 , -1.76] & [-3.40 , -1.54] & [-3.90 , -1.51] \\
        $\log_{10}(\phi_{0,\textrm{Sp}}^*)$ & -2.04 & [-2.08 , -1.99] & [-2.14 , -1.94] & [-2.20 , -1.91] \\
    \end{tabular}
\end{center}
{\it Note:} parameter values are given for $H_0=100$ kms$^{-1}$Mpc$^{-1}$ and for $z=0$ ($\phi^*$ is in units of Mpc$^{-3}$mag$^{-1}$).
\end{table*}

\begin{table*}
\caption{1D confidence intervals for the five insets of the D1 observed data and the joint data.}
\label{table:app2_real}
\begin{center}
    \begin{tabular}{c c c c}
        \hline\hline
         & 68\% & 95\% & 99\% \\
        \hline
        \text{real 1} & & & \\
        $\log_{10}(\phi_{0,\textrm{E}}^*)$ & [-1.75 , -1.47] & [-1.94 , -1.35] & [-2.12 , -1.29] \\
        $\log_{10}(\phi_{0,\textrm{Sp}}^*)$ & [-2.08 , -1.95] & [-2.17 , -1.89] & [-2.31 , -1.83] \\
        \hline
        \text{real 2} & & & \\
        $\log_{10}(\phi_{0,\textrm{E}}^*)$ & [-2.01 , -1.60] & [-2.31 , -1.43] & [-2.89 , -1.26] \\
        $\log_{10}(\phi_{0,\textrm{Sp}}^*)$ & [-2.08 , -1.93] & [-2.17 , -1.85] & [-2.27 , -1.53] \\
        \hline
        \text{real 3} & & & \\
        $\log_{10}(\phi_{0,\textrm{E}}^*)$ & [-1.92 , -1.58] & [-2.16 , -1.47] & [-2.51 , -1.33] \\
        $\log_{10}(\phi_{0,\textrm{Sp}}^*)$ & [-2.17 , -2.02] & [-2.24 , -1.96] & [-2.46 , -1.96] \\
        \hline
        \text{real 4} & & & \\
        $\log_{10}(\phi_{0,\textrm{E}}^*)$ & [-2.03 , -1.63] & [-2.30 , -1.47] & [-2.76 , -1.31] \\
        $\log_{10}(\phi_{0,\textrm{Sp}}^*)$ & [-2.06 , -1.94] & [-2.15 , -1.89] & [-2.22 , -1.84] \\
        \hline
        \text{real 5} & & & \\
        $\log_{10}(\phi_{0,\textrm{E}}^*)$ & [-1.93 , -1.54] & [-2.28 , -1.34] & [-2.90 , -1.24] \\
        $\log_{10}(\phi_{0,\textrm{Sp}}^*)$ & [-2.13 , -1.96] & [-2.23 , -1.88] & [-2.44 , -1.73] \\
        \hline
        \text{chain-rule joint} & & & \\
        $\log_{10}(\phi_{0,\textrm{E}}^*)$ & [-1.84 , -1.58] & [-2.05 , -1.48] & [-2.39 , -1.38] \\
        $\log_{10}(\phi_{0,\textrm{Sp}}^*)$ & [-2.10 , -2.00] & [-2.15 , -1.96] & [-2.23 , -1.92] \\
        \hline
    \end{tabular}
\end{center}
{\it Note:} parameter values are given for $H_0=100$ kms$^{-1}$Mpc$^{-1}$ and for $z=0$ ($\phi^*$ is in units of Mpc$^{-3}$mag$^{-1}$).
\end{table*}

We emphasize that for the observed and virtual data, Figs. \ref{fig:app2_PMC_fake} and  \ref{fig:app2_PMC_real} show that the results are narrower 1D marginalized distributions for the spiral population than for the elliptical population. This comes from the steep faint-end slope of the luminosity function for the spirals: a small change in $\log_{10}(\phi_{0,\textrm{Sp}}^*)$ has a strong effect on the number of faint spirals, which can be more easily distinguished. This is not exactly the case for the other parameters: a strong change in $\log_{10}(\phi_{0,\textrm{E}}^*)$ is needed to considerably alter the number of elliptical galaxies, which equates to greater uncertainty on this parameter.

Comparison of Figs. \ref{fig:app2_PMC_fake} and \ref{fig:app2_PMC_real} show that if the posterior distributions of the spirals have a similar dispersion for the virtual and observed data, the posterior distributions of the elliptical population is narrower for the observed than for the virtual data. This is an effect of the small number of ellipticals in an image: with a low value for $\log_{10}(\phi_{0,E}^*)$, there are very few elliptical galaxies in the image, and this lack of a statistically significant sample increases the width of the posterior due to the lack of available information. In comparison, the posterior is narrower for the observed data because more elliptical galaxies appear to be present: the peak of the joint posterior is at $\log_{10}(\phi_{0,E}^*)=-1.7$, compared to a lower value of $\log_{10}(\phi_{0,E}^*)=-2.1$ for the virtual data (see Tables \ref{table:app2_fake} and \ref{table:app2_real}). The summaries provided by the network encode information about the relation of the number of ellipticals and spirals in the images to the model parameters, and therefore can be used for the inference of this property using the PMC.

For the observed data frame labeled "real 3" in Fig. \ref{fig:app2_PMC_real}, the 1D confidence interval of the parameter $\log_{10}(\phi_{0,Sp}^*)$ is slightly displaced compared to the other observed images, and moreover toward lower densities of spirals. This can be explained by the presence of a large spiral galaxy in the image of frame "c" in Fig. \ref{fig:app2_D1}. This galaxy covers $\sim 100\times100\; \textrm{pixels}$, an area of the image (i.e., $\sim1\%$ of the full $1024\times1024$ image) that reduces the amount of informative data with which to correlate the number of detected spirals, which therefore appear to be fewer because the area in which they are visible is smaller. This illustrates the fact that the inference is correct even if the PMC procedure happens to be biased because of statistically anomalous data. This is the reason why we used five insets of the D1 deep field, with the goal to improve the robustness of the results.

\subsection{Joint posterior and confidence intervals}\label{sec:joint-pmc}

Because we only had individual posteriors for each image, we combined to derive a unique posterior. Unfortunately, posterior chains cannot be combined in a simple way. A rigorous way to achieve such a posterior is to use the Bayesian chain rule,
\begin{equation}
\begin{array}{ccl}
    p(\theta|X_1,\dots,X_n) & \propto & p(\theta)p(X_1|\theta)p(X_2|X_1,\theta)\cdots \\
     & & p(X_n|X_1,\dots,X_{n-1},\theta),
\end{array}
\label{eq:joint-chain}
\end{equation}
where $\theta$ is a set of parameters, and $X_1,\dots,X_n$ are the $n$ individual pieces of data. The chain rule allows obtaining the posteriors sequentially: $\forall i\in\{1,\dots,n-1\}$ we assumed that obtained the posterior distribution (via PMC) of $p(\theta|X_1,\dots,X_i)$, then 
\begin{enumerate}
    \item Consider $p(\theta|X_1,\dots,X_i)$, derived from the pieces of data $X_1,\dots,X_i$, as an approximate proposal distribution for $\theta$.
    \item Use that proposal distribution of $\theta$ for the inference, using ABC or PMC, given the new piece of data, $X_{i+1}$.
    \item Derive a new posterior distribution from $p(\theta|X_1,\dots,X_{i+1})$ that can be used in turn as the proposal distribution for the next piece of data.
\end{enumerate}

We ran a parallelized version of such a sequential-PMC for the five images for the virtual and observed data. The resulting joint posteriors are shown as black lines (called kernel density estimate ''KDE joint'') in Figs. \ref{fig:app2_PMC_fake} and \ref{fig:app2_PMC_real} and can be compared to the posterior distributions obtained for the individual images. In these figures, the $68\%$, $95\%$ and $99\%$ contours are drawn using the joint posterior in black, gray, and light gray, respectively. The 1D  $68\%$, $95\%,$ and $99\%$ confidence intervals for the joint posteriors are listed at the end of Table \ref{table:app2_fake} for the virtual data and in Table \ref{table:app2_real} for the observed data. The confidence intervals are narrower for the joint posteriors because more information is available, which allows us to draw tighter constraints on the parameter values.

\subsection{Comparison with other studies}

We compared our results with other measurements of the luminosity functions for the elliptical (or red, or quiescent) galaxy population and for the spiral (or blue, or star-forming) population obtained from deep multiband galaxy surveys: \citet{2017A&A...599A..62L}, \citet{2007ApJ...654..858B}, \citet{2015ApJ...815...94B}, \citet{2008A&A...477..763S}, \citet{2009ApJ...707.1595D}, and \citet{2007ApJ...665..265F}. In these studies, the two populations are identified either by comparing the observed magnitudes in different bands or by finding the best-fit SEDs to each galaxy. These methods are affected by several biases because of the catalog extraction process (see Sect. \ref{sec:intro}) and because of the choice of threshold used to split galaxies into red or blue.

\begin{figure}
    \includegraphics[width=0.5\textwidth]{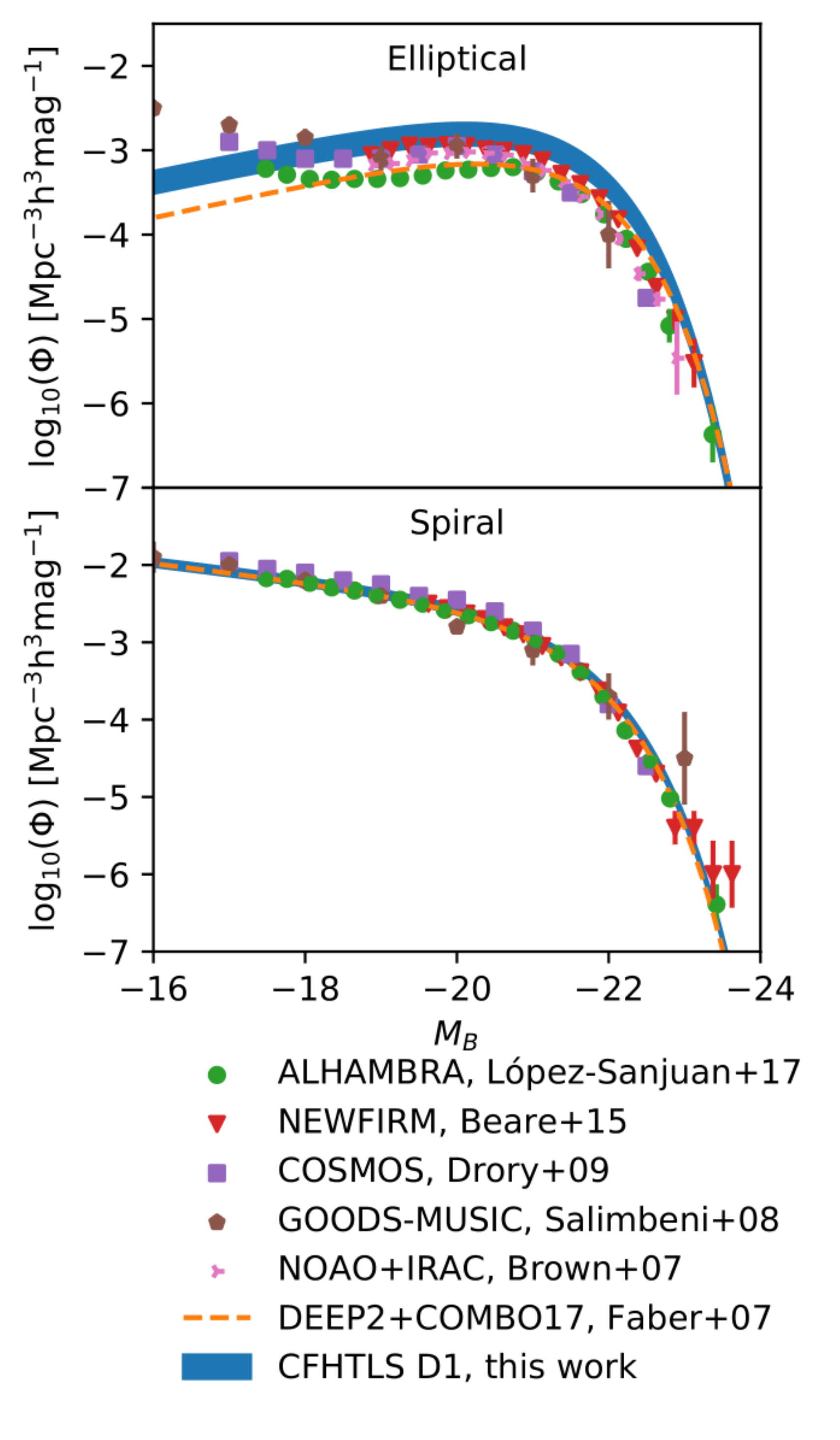}
    \caption{Inferred luminosity functions derived for the elliptical (top) and the spiral (bottom) populations using the $68\%$ confidence intervals (blue). Our results are compared with the results of \citet{2017A&A...599A..62L} in green, \citet{2007ApJ...654..858B} in pink, \citet{2015ApJ...815...94B} in red, \citet{2008A&A...477..763S} in brown, \citet{2009ApJ...707.1595D} in purple, and \citet{2007ApJ...665..265F} in orange. Our $g'$ magnitudes are converted into the Johnson B-band absolute magnitude, at redshift $z=0.5$ and for $H_0=100$ kms$^{-1}$Mpc$^{-1}$ using $B-g'=0.78$ for ellipticals and $B-g'=0.40$ for spirals, listed in Table 7 of \citet{1995PASP..107..945F}.}
    \label{fig:LF_comparison}
\end{figure}

Fig. \ref{fig:LF_comparison} shows the luminosity function using the $68\%$ confidence intervals of the joint posterior over the five insets of the CFHTLS+WIRDS D1 field in blue for the elliptical (top) and spiral populations (bottom) compared to the results from the other articles. Fig. \ref{fig:LF_comparison} shows a good agreement between our results and the other surveys for the spiral population (bottom panel) for which the network was able to precisely constrain the luminosity function (narrow $68\%$ confidence interval). Despite a different sample and no catalog extraction, when our results are compared to those of \citet{2017A&A...599A..62L}, we obtain for the dominant spiral population a 1$\sigma$ confidence interval on $\phi^*$ comparable to theirs. The top panel of Fig. \ref{fig:LF_comparison} shows that for the elliptical population, there are larger differences between all analyses, including ours. This is likely partly due to Poisson noise:  the number density of galaxies in the elliptical population is typically a factor of 10 to 100 lower than for the spiral population in a given magnitude bin, introducing more dispersion. Moreover, as pointed out in several of the analyses we quoted, the chosen samples of red or blue galaxies contain misidentified galaxies, which affect the smaller elliptical population more.

The top panel of Fig. \ref{fig:LF_comparison} also indicates evidence for an excess of red faint galaxies: \citet{2017A&A...599A..62L}, \citet{2009ApJ...707.1595D}, and \citet{2008A&A...477..763S} have used the sum of two Schechter functions to better model the luminosity function of the elliptical population (and its rising faint end). We suspect that this apparently observed excess of faint elliptical galaxies to which our method is in principle quite sensitive to tends to pull our single-Schechter luminosity function upward for the elliptical population and is therefore systematically higher than the luminosity functions derived by \citet{2007ApJ...654..858B}, \citet{2015ApJ...815...94B}, and \citet{2007ApJ...665..265F}.
In this case, our limited model of two galaxy populations with single evolving Schechter luminosity functions for each population is too coarse. Nevertheless, the rising of the elliptical luminosity function at the faint end could be caused by some of the faint spiral galaxies, which are numerous, being categorized as red galaxies by some source extraction method.

Conversely, some of the elliptical galaxies in the other analyses may have been misclassified by some source extraction method as blue galaxies and might lead to a systematically too low amplitude for the elliptical luminosity function. Moreover, dusty star-forming galaxies (which have red colors) may be modeled as ellipticals in our analysis (due to the resemblance of the strong starburst to a bulge-dominated object in the image), whereas they may be classified as spirals in the other analyses. Both effects would affect only the luminosity function of elliptical galaxies because spiral types vastly dominate the elliptical types in number.

\section{Conclusions and perspectives}

We have introduced a novel method to infer robust constraints on the luminosity functions of elliptical and spiral galaxies using a massive compression of panchromatic images through a neural network and likelihood-free (simulation-based) inference. This approach directly analyzes multiband deep-field pixel maps using neural networks and performs a likelihood-free inference without the need for any source catalog. The use of simulated images in similar ``observing'' conditions as the data, taking the complex selection biases that affect the survey images into account (see Sect. \ref{sec:intro}), as a central part of the inference process and the direct comparison of the simulations to the observed CFHTLS deep field is the key contribution of our approach. 

In this article, synthetic populations of elliptical and spiral galaxies were simulated using a forward model of galaxy evolution, sampled from luminosity functions for each population and decomposed into their bulge and disk through a set of SEDs. The forward model was made even more realistic by adding the internal extinction by dust of each component of the galaxies, the reddening caused by the Milky Way, and the stars from the Besançon stellar model. The \texttt{SkyMaker} software was then used to paint these simulated galaxies and stars in realistic panchromatic images in the optical $u'$, $g'$, $r'$, $i'$, $z'$ MegaPrime filters and the near-IR $J$, $H$, $Ks$  WIRCam  filters, making sure to reproduce the instrumental selection effects of CFHTLS and WIRCAM images (the parameters of the forward model are summarized in Table \ref{table:overview}).\\
\\
The images simulated through this process were then used to train a fully convolutional inception network to extract information about the parameters of each luminosity function. The network is trained to maximize the Fisher information of the images and enables the drastic reduction in the dimensionality of the images down to the number of parameters of the model. In contrast to deep-learning approaches, training the neural network with only 400 images (200 for the fiducial parameters, and 200 for the derivatives) is sufficient to obtain very good results. After the network was trained on simulated deep fields, ABC/PMC procedures were run, starting from a uniform prior, to infer the parameters of the luminosity functions of data images. The approach was applied to virtual data and insets of the observed D1 deep field. We showed that we can robustly infer the parameters used to generate the virtual data: $\log_{10}(\phi^*)$ and $M^*$ for a spiral population in Sect. \ref{sec:app1}, and $\log_{10}(\phi^*)$ for spiral and elliptical galaxies in Sect. \ref{sec:app2}. We also developed a joint-PMC procedure in order to infer the parameters using multiple $1024\times 1024$ pixel images at once.\\
\\
This method proved its efficiency in constraining the nontrivial and correlated parameters of two luminosity functions of elliptical and spiral populations of galaxies: the amplitude and the characteristic magnitude. Using the likelihood-free inference and the compressed network summaries, we were able to infer possible input values of the parameters for virtual data as well as the values of the model parameters describing the generative process for insets of the observed CFHTLS D1 data. The derived luminosity functions agree well with those for the elliptical (or red, or quiescent) galaxy population and for the spiral (or blue, or star-forming) population obtained from deep multiband galaxy surveys by \citet{2017A&A...599A..62L}, \citet{2007ApJ...654..858B}, \citet{2015ApJ...815...94B}, \citet{2008A&A...477..763S}, \citet{2009ApJ...707.1595D}, and \citet{2007ApJ...665..265F}, especially for the spiral population. For the elliptical population, information about the excess of faint galaxies that some authors have tried to model as the sum of two Schechter functions was encoded via the IMNN and yields a higher amplitude for this galaxy type.

We have illustrated the method using only two parameters of the luminosity functions of two galaxy types, but in principle, the usual five parameters of evolving luminosity functions (amplitude, characteristic magnitude, faint-end slope, amplitude evolution, and characteristic magnitude evolution) could be simultaneously inferred from observed data. Further realism in the simulations is still required, including all galaxy types that significantly contribute to the observed deep fields, such as lenticular galaxies; it is not clear whether a population of irregular galaxies should be considered; but dividing the spirals galaxies into early and late type spirals (with a high bulge-to-total ratio for Sa-Sb types and a low ratio for Sc-Sd types) may also bring useful information if these types evolve differently from redshift $z=1$.

In the model used here, a galaxy was decomposed into a bulge and a disk, which is too simplistic for dusty galaxies. F. Livet is currently refining the model by decomposing the disk into a thin and a thick disk to better model the color gradients of spiral galaxies. Our current forward model uses two nonevolving SEDs of \citet{1980ApJS...43..393C} to model the bulges and disks of spiral galaxies, but it can be improved with evolving scenarios of galaxies, and their bulge and disk, such as those of the Pegase model of \citet{2019A&A...623A.143F}.

In any case, the IMNN compresses the input to the same dimension as the parameter space, and this extreme compression enables us to potentially investigate a large number of physical parameters. The limiting step would be that the ABC or PMC procedure might lead to a very long exploration of the parameters space due to its high dimensionality, but we could use the \texttt{pydelfi} approach of \citet{2019MNRAS.488.4440A} to explore other likelihood-free inference techniques that are less expensive in terms of the number of required simulations. The computing time clearly is the limiting factor for a realistic multipopulation analysis of one full CFHTLS deep field.

\section*{Acknowledgments}

Florian Livet upgraded and optimized the code for the generation of the forward simulations, contributed to the development of the IMNN, wrote and optimized the inception architecture, developed a Docker environment to parallelize the PMC and to apply the Bayesian chain rule. Tom Charnock led the code development for the IMNN, provided practical and theoretical expertise to all the statistical analyses and methods involved in the IMNN and PMC and technical and methodological concepts. Valérie de Lapparent and Damien Le Borgne are the supervisors of the PhD of Florian Livet as experts on the astrophysical and statistical aspects of the analysis. Damien Le Borgne provided his coding expertise and compared the observed and predicted galaxy/stellar number counts.

The authors thank Emmanuel Bertin for managing the \texttt{Morpho} cluster used to generate all the simulations and perform all the network trainings, for his available programs (\texttt{Stuff} and \texttt{SkyMaker}) and for his help to create a \texttt{Docker} environment. This work uses the available final releases of the TERAPIX processed data, T0007 for CFHTLS  \citep{2012yCat.2317....0H} and T0002 for WIRDS \citep{2012A&A...545A..23B}. Our forward model was extended using the stellar distributions of the Besançon model \citep{2003A&A...409..523R} web service. Tom Charnock acknowledges financial support from the Sorbonne Univ. Emergence fund, 2019-2020.


\begin{table*}
\caption{Parameters for the \texttt{SkyMaker} program}
\label{table:skymaker}
\begin{center}
    \begin{tabular}{l c c c c c c c c}
        \hline\hline
        Passband & $u$ & $g$ & $r$ & $i$ & $z$ & $J$ & $H$ & $K\text{s}$\\
        \hline
        Effective wavelength ($\mu$m) & 0.355 & 0.475 & 0.640 & 0.776 & 0.925 & 1.253 & 1.631 & 2.146 \\
        Effective gain (e$^-$/ADU) & 1026 & 7095 & 6919 & 6807 & 2148 & 1116 & 2379 & 2134 \\
        Effective exposure time (sec) & 1 & 1 & 1 & 1 & 1 & 1 & 1 & 1 \\
        Saturation level (ADU) & 6466 & 2710 & 3088 & 4231 & 12331 & 68068 & 114324 & 110884 \\
        Read-out noise (e$^-$) & 5 & 5 & 5 & 5 & 5 & 30 & 30 & 30 \\
        Seeing FWHM (arcsec) & 0.833 & 0.817 & 0.771 & 0.744 & 0.718 & 0.65 & 0.65 & 0.65 \\
        Sky level (AB mag/arcsec$^2$) & 22.2 & 21.7 & 20.8 & 20.0 & 19.1 & 16.7 & 15.4 & 15.4 \\
        Zero-point (AB mag) & 30 & 30 & 30 & 30 & 30 & 30 & 30 & 30 \\
        Milky Way reddening (AB mag) & 0.097 & 0.075 & 0.049 & 0.036 & 0.027 & 0.016 & 0.010 & 0.007 \\
        \hline
    \end{tabular}
\end{center}
\end{table*}

\begin{table*}
\caption{Description of the network (see Fig. \ref{fig:architecture}) used in Sects. \ref{sec:app1} and \ref{sec:app2}.}
\label{table:architecture}
\begin{center}
    \small
    \begin{tabular}{c c c c c c c}
        \hline\hline
        Label & Width & Height & Channels & Stride & Padding & Activation \\
        \hline
        Input & 1024 & 1024 & 8 & $\emptyset$ & $\emptyset$ & $\emptyset$ \\
        $1\times 1$ convolution & 1 & 1 & 8 & 1 & SAME & Leaky ReLU 0.01 \\
        $3\times 1$ convolution & 3 & 1 & 8 & 1 & SAME & Leaky ReLU 0.01 \\
        $1\times 3$ convolution & 1 & 3 & 8 & 1 & SAME & Leaky ReLU 0.01 \\
        $3\times 3$ convolution & \multicolumn{6}{c}{$3\times 1$ convolution and $1\times 3$ convolution concatenated} \\
        $5\times 1$ convolution & 5 & 1 & 8 & 1 & SAME & Leaky ReLU 0.01 \\
        $1\times 5$ convolution & 1 & 5 & 8 & 1 & SAME & Leaky ReLU 0.01 \\
        $5\times 5$ convolution & \multicolumn{6}{c}{$5\times 1$ convolution and $1\times 5$ convolution concatenated} \\
        $2\times 2$ max pooling & 2 & 2 & $\emptyset$ & 2 & $\emptyset$ & $\emptyset$ \\
        $4\times 4$ avg pooling & 4 & 4 & $\emptyset$ & 4 & $\emptyset$ & $\emptyset$ \\
        Output & 1 & 1 & 2 & $\emptyset$ & $\emptyset$ & Linear \\
        \hline
    \end{tabular}
\end{center}
{\it Note:} Each line of this table describes the boxes of Fig. \ref{fig:architecture}.
\end{table*}

\begin{table*}
\caption{Optimizer and learning rate used during the training of Sect. \ref{sec:app1}.}
\label{table:app1_lr}
\begin{center}
    \small
    \begin{tabular}{c c c}
        \hline\hline
        epoch & optimizer & learning rate \\
        \hline
        0 & Adam & 0.02 \\
        100 & SGD & $1\times 10^{-6}$ \\
        200 & SGD & $3\times 10^{-6}$ \\
        600 & SGD & $2\times 10^{-6}$ \\
        6300 & SGD & $1.5\times 10^{-6}$ \\
        8000 & SGD & $1\times 10^{-6}$ \\
        8500 & SGD & $7.5\times 10^{-7}$ \\
        13300 & SGD & $5.5\times 10^{-7}$ \\
        \hline
    \end{tabular}
\end{center}
{\it Note:} SGD stands for stochastic gradient descent.
\end{table*}

\begin{table*}
\caption{Overview of the parameters of the forward model for Sects. \ref{sec:app1} and \ref{sec:app2}.}
\label{table:overview}
\begin{center}
    \begin{small}
    \begin{tabular}{c c c}
         & Application Sect. \ref{sec:app1} & Application Sect. \ref{sec:app2} \\
        \hline\hline
        COSMOLOGY & & \\
        Hubble constant & \multicolumn{2}{c}{$H_0=70$ km s$^{-1}$ Mpc$^{-1}$} \\
        Matter density & \multicolumn{2}{c}{$\Omega_M=0.3$} \\
        Cosmological constant & \multicolumn{2}{c}{$\Omega_\Lambda=0.7$} \\
        Redshift integration step & \multicolumn{2}{c}{$5$ h$^{-1}$ Mpc} \\
        \hline
        IMAGE & & \\
        Pixel size & \multicolumn{2}{c}{$0.186$ arcsec} \\
        Field size & \multicolumn{2}{c}{$1024\times 1024$ px} \\
        Apparent magnitude limits & \multicolumn{2}{c}{$[17, 29]$ for SDSS $g'$ filter} \\
        \hline
        ZERO POINTS / BACKGROUND & & \\
        Collecting area & \multicolumn{2}{c}{$10.17$} \\
        Gain & \multicolumn{2}{c}{$1.67$} \\
        Background SED & \multicolumn{2}{c}{puxley} \\
        \hline
        PASSBANDS & & \\
        Reference & \multicolumn{2}{c}{SDSS $g'$} \\
        Observed & \multicolumn{2}{c}{MegaPrime $u$, $g$, $r$, $i$, $z$, WIRCam $J$, $H$, $Ks$} \\
        Detection type & \multicolumn{2}{c}{Photons} \\
        Calibration & \multicolumn{2}{c}{AB} \\
        \hline
        ELLIPTICAL POPULATION & & \\
        Hubble type & & $-6$ \\
        Bulge to total & & $1$ \\
        Bulge SED$^1$ & & E \\
        Disk SED$^1$ & & $\emptyset$ \\
        $log_{10}(\phi_0^*)$ & & $-2.09$ Mpc$^{-3}$ mag$^{-1}$ (fiducial) \\
        $M_0^*$ & & $-19.68$ \\
        $\alpha$ & & $-0.53$ \\
        $\phi_\text{evol}^*$ & & $-1.37$ \\
        $M_\text{evol}^*$ & & $-1.15$ \\
        \hline
        SPIRAL POPULATION & & \\
        Hubble type & $4$ & $4$ \\
        Bulge to total & $0.2$ & $0.2$ \\
        Bulge SED$^1$ & E & E \\
        Disk SED$^1$ & Irr & Irr \\
        $log_{10}(\phi_0^*)$ & $-2.01$ Mpc$^{-3}$ mag$^{-1}$ (fiducial) & $-2.04$ Mpc$^{-3}$ mag$^{-1}$ (fiducial) \\
        $M_0^*$ & $-20.0$ (fiducial) & $-19.71$ \\
        $\alpha$ & $-1.3$ & $-1.29$ \\
        $\phi_\text{evol}^*$ & $0$ & $-0.03$ \\
        $M_\text{evol}^*$ & $0$ & $-1.49$ \\
        \hline
        EXTINCTION & & \\
        Extinction curve & \multicolumn{2}{c}{Milky Way$^2$} \\
        Reddening & \multicolumn{2}{c}{0.097 ($u$), 0.0745 ($g$), 0.049 ($r$), 0.036 ($i$),} \\
         & \multicolumn{2}{c}{0.027 ($z$), 0.016 ($J$), 0.010 ($H$), 0.007 ($Ks$)} \\
        Inter-galactic medium & \multicolumn{2}{c}{Madau model$^3$} \\
        \hline
        STARS & \multicolumn{2}{c}{Besançon model$^4$} \\
        \hline
        BULGE PARAMETERS$^5$ & & \\
        $R_0$ & \multicolumn{2}{c}{$1.58$ $h^{-1}$ kpc} \\
        $M_0$ & \multicolumn{2}{c}{$-20.5$} \\
        $\gamma_B$ & \multicolumn{2}{c}{$-1$} \\
        \hline
        DISK PARAMETERS$^6$ & & \\
        $h_0$ & \multicolumn{2}{c}{$3.85$ $h^{-1}$ kpc} \\
        $\beta$ & \multicolumn{2}{c}{$-0.214$} \\
        $\sigma$ & \multicolumn{2}{c}{$0.36$} \\
        $\gamma_D$ & \multicolumn{2}{c}{$-0.8$} \\
        \hline
    \end{tabular}
    \end{small}
\end{center}
\begin{list}{}{}
\small
\item {\it Notes:}
\item $^1$ The SED templates are taken from \citet{1980ApJS...43..393C} for the far UV and visible and extended to the near-IR.
\item $^2$ The extinction curve is taken from \citet{Fitzpatricketal_2007} and extended to the near-IR.
\item $^3$ The extinction curve is corrected for the intergalactic medium effects as in \citet{1996MNRAS.283.1388M}.
\item $^4$ Stars are added following the Besançon model of \citet{2003A&A...409..523R} and their recent improvements (\citet{2012A&A...538A.106R, 2014A&A...569A..13R,  2015A&A...581A.123B, 2017A&A...602A..67A}).
\item $^5$ For a full description of these parameters, see Sect. \ref{sec:bulge}.
\item $^6$ For a full description of these parameters see Sect. \ref{sec:disk}.
\end{list}
\end{table*}

\bibliography{biblio}

\begin{thebibliography}{68}
\expandafter\ifx\csname natexlab\endcsname\relax\def\natexlab#1{#1}\fi

\bibitem[{{Abazajian} {et~al.}(2009){Abazajian}, {Adelman-McCarthy},
  {Ag{\"u}eros}, {Allam}, {Allende Prieto}, {An}, {Anderson}, {Anderson},
  {Annis}, {Bahcall}, {Bailer-Jones}, {Barentine}, {Bassett}, {Becker},
  {Beers}, {Bell}, {Belokurov}, {Berlind}, {Berman}, {Bernardi}, {Bickerton},
  {Bizyaev}, {Blakeslee}, {Blanton}, {Bochanski}, {Boroski}, {Brewington},
  {Brinchmann}, {Brinkmann}, {Brunner}, {Budav{\'a}ri}, {Carey}, {Carliles},
  {Carr}, {Castander}, {Cinabro}, {Connolly}, {Csabai}, {Cunha}, {Czarapata},
  {Davenport}, {de Haas}, {Dilday}, {Doi}, {Eisenstein}, {Evans}, {Evans},
  {Fan}, {Friedman}, {Frieman}, {Fukugita}, {G{\"a}nsicke}, {Gates},
  {Gillespie}, {Gilmore}, {Gonzalez}, {Gonzalez}, {Grebel}, {Gunn},
  {Gy{\"o}ry}, {Hall}, {Harding}, {Harris}, {Harvanek}, {Hawley}, {Hayes},
  {Heckman}, {Hendry}, {Hennessy}, {Hindsley}, {Hoblitt}, {Hogan}, {Hogg},
  {Holtzman}, {Hyde}, {Ichikawa}, {Ichikawa}, {Im}, {Ivezi{\'c}}, {Jester},
  {Jiang}, {Johnson}, {Jorgensen}, {Juri{\'c}}, {Kent}, {Kessler}, {Kleinman},
  {Knapp}, {Konishi}, {Kron}, {Krzesinski}, {Kuropatkin}, {Lampeitl},
  {Lebedeva}, {Lee}, {Lee}, {French Leger}, {L{\'e}pine}, {Li}, {Lima}, {Lin},
  {Long}, {Loomis}, {Loveday}, {Lupton}, {Magnier}, {Malanushenko},
  {Malanushenko}, {Mand elbaum}, {Margon}, {Marriner}, {Mart{\'\i}nez-Delgado},
  {Matsubara}, {McGehee}, {McKay}, {Meiksin}, {Morrison}, {Mullally}, {Munn},
  {Murphy}, {Nash}, {Nebot}, {Neilsen}, {Newberg}, {Newman}, {Nichol},
  {Nicinski}, {Nieto-Santisteban}, {Nitta}, {Okamura}, {Oravetz}, {Ostriker},
  {Owen}, {Padmanabhan}, {Pan}, {Park}, {Pauls}, {Peoples}, {Percival}, {Pier},
  {Pope}, {Pourbaix}, {Price}, {Purger}, {Quinn}, {Raddick}, {Re Fiorentin},
  {Richards}, {Richmond}, {Riess}, {Rix}, {Rockosi}, {Sako}, {Schlegel},
  {Schneider}, {Scholz}, {Schreiber}, {Schwope}, {Seljak}, {Sesar}, {Sheldon},
  {Shimasaku}, {Sibley}, {Simmons}, {Sivarani}, {Allyn Smith}, {Smith},
  {Smol{\v{c}}i{\'c}}, {Snedden}, {Stebbins}, {Steinmetz}, {Stoughton},
  {Strauss}, {SubbaRao}, {Suto}, {Szalay}, {Szapudi}, {Szkody}, {Tanaka},
  {Tegmark}, {Teodoro}, {Thakar}, {Tremonti}, {Tucker}, {Uomoto}, {Vanden
  Berk}, {Vandenberg}, {Vidrih}, {Vogeley}, {Voges}, {Vogt}, {Wadadekar},
  {Watters}, {Weinberg}, {West}, {White}, {Wilhite}, {Wonders}, {Yanny},
  {Yocum}, {York}, {Zehavi}, {Zibetti}, \& {Zucker}}]{2009ApJS..182..543A}
{Abazajian}, K.~N., {Adelman-McCarthy}, J.~K., {Ag{\"u}eros}, M.~A., {et~al.}
  2009, \apjs, 182, 543

\bibitem[{{Akeret} {et~al.}(2015){Akeret}, {Refregier}, {Amara}, {Seehars}, \&
  {Hasner}}]{2015JCAP...08..043A}
{Akeret}, J., {Refregier}, A., {Amara}, A., {Seehars}, S., \& {Hasner}, C.
  2015, \jcap, 2015, 043

\bibitem[{{Alsing} {et~al.}(2019){Alsing}, {Charnock}, {Feeney}, \&
  {Wandelt}}]{2019MNRAS.488.4440A}
{Alsing}, J., {Charnock}, T., {Feeney}, S., \& {Wandelt}, B. 2019, \mnras, 488,
  4440

\bibitem[{{Alsing} \& {Wandelt}(2018)}]{2018MNRAS.476L..60A}
{Alsing}, J. \& {Wandelt}, B. 2018, \mnras, 476, L60

\bibitem[{{Am{\^o}res} {et~al.}(2017){Am{\^o}res}, {Robin}, \&
  {Reyl{\'e}}}]{2017A&A...602A..67A}
{Am{\^o}res}, E.~B., {Robin}, A.~C., \& {Reyl{\'e}}, C. 2017, \aap, 602, A67

\bibitem[{{Beare} {et~al.}(2015){Beare}, {Brown}, {Pimbblet}, {Bian}, \&
  {Lin}}]{2015ApJ...815...94B}
{Beare}, R., {Brown}, M. J.~I., {Pimbblet}, K., {Bian}, F., \& {Lin}, Y.-T.
  2015, \apj, 815, 94

\bibitem[{{Bernardi} {et~al.}(2017){Bernardi}, {Fischer}, {Sheth}, {Meert},
  {Huertas-Company}, {Shankar}, \& {Vikram}}]{2017MNRAS.468.2569B}
{Bernardi}, M., {Fischer}, J.~L., {Sheth}, R.~K., {et~al.} 2017, \mnras, 468,
  2569

\bibitem[{{Bertin}(2009)}]{2009MmSAI..80..422B}
{Bertin}, E. 2009, \memsai, 80, 422

\bibitem[{{Bertin}(2010)}]{2010ascl.soft10067B}
{Bertin}, E. 2010, {Stuff: Simulating ``Perfect'' Astronomical Catalogues}

\bibitem[{{Bertin}(2011)}]{2011ASPC..442..435B}
{Bertin}, E. 2011, Astronomical Society of the Pacific Conference Series, Vol.
  442, {Automated Morphometry with SExtractor and PSFEx}, ed. I.~N. {Evans},
  A.~{Accomazzi}, D.~J. {Mink}, \& A.~H. {Rots}, 435

\bibitem[{{Bertin} \& {Arnouts}(1996)}]{1996A&AS..117..393B}
{Bertin}, E. \& {Arnouts}, S. 1996, \aaps, 117, 393

\bibitem[{{Bertin} \& {Arnouts}(2010)}]{2010ascl.soft10064B}
{Bertin}, E. \& {Arnouts}, S. 2010, {SExtractor: Source Extractor}

\bibitem[{{Bielby} {et~al.}(2012){Bielby}, {Hudelot}, {McCracken}, {Ilbert},
  {Daddi}, {Le F{\`e}vre}, {Gonzalez-Perez}, {Kneib}, {Marmo}, {Mellier},
  {Salvato}, {Sanders}, \& {Willott}}]{2012A&A...545A..23B}
{Bielby}, R., {Hudelot}, P., {McCracken}, H.~J., {et~al.} 2012, \aap, 545, A23

\bibitem[{{Bienaym{\'e}} {et~al.}(2015){Bienaym{\'e}}, {Robin}, \&
  {Famaey}}]{2015A&A...581A.123B}
{Bienaym{\'e}}, O., {Robin}, A.~C., \& {Famaey}, B. 2015, \aap, 581, A123

\bibitem[{{Binggeli} {et~al.}(1984){Binggeli}, {Sandage}, \&
  {Tarenghi}}]{1984AJ.....89...64B}
{Binggeli}, B., {Sandage}, A., \& {Tarenghi}, M. 1984, \aj, 89, 64

\bibitem[{{Brown} {et~al.}(2007){Brown}, {Dey}, {Jannuzi}, {Brand }, {Benson},
  {Brodwin}, {Croton}, \& {Eisenhardt}}]{2007ApJ...654..858B}
{Brown}, M. J.~I., {Dey}, A., {Jannuzi}, B.~T., {et~al.} 2007, \apj, 654, 858

\bibitem[{{Calvi} {et~al.}(2014){Calvi}, {Stiavelli}, {Bradley}, {Pizzella}, \&
  {Kim}}]{2014ApJ...796..102C}
{Calvi}, V., {Stiavelli}, M., {Bradley}, L., {Pizzella}, A., \& {Kim}, S. 2014,
  \apj, 796, 102

\bibitem[{{Carassou} {et~al.}(2017){Carassou}, {de Lapparent}, {Bertin}, \& {Le
  Borgne}}]{2017A&A...605A...9C}
{Carassou}, S., {de Lapparent}, V., {Bertin}, E., \& {Le Borgne}, D. 2017,
  \aap, 605, A9

\bibitem[{{Charnock} {et~al.}(2018){Charnock}, {Lavaux}, \&
  {Wandelt}}]{2018PhRvD..97h3004C}
{Charnock}, T., {Lavaux}, G., \& {Wandelt}, B.~D. 2018, \prd, 97, 083004

\bibitem[{{Chevallard} {et~al.}(2013){Chevallard}, {Charlot}, {Wandelt}, \&
  {Wild}}]{2013MNRAS.432.2061C}
{Chevallard}, J., {Charlot}, S., {Wandelt}, B., \& {Wild}, V. 2013, \mnras,
  432, 2061

\bibitem[{Cireşan {et~al.}(2010)Cireşan, Meier, Gambardella, \&
  Schmidhuber}]{CIRESAN2010}
Cireşan, D.~C., Meier, U., Gambardella, L.~M., \& Schmidhuber, J. 2010, Neural
  Computation, 22, 3207, pMID: 20858131

\bibitem[{{Cisewski-Kehe} {et~al.}(2019){Cisewski-Kehe}, {Weller}, \&
  {Schafer}}]{2019arXiv190411306C}
{Cisewski-Kehe}, J., {Weller}, G., \& {Schafer}, C. 2019, arXiv e-prints,
  arXiv:1904.11306

\bibitem[{{Coleman} {et~al.}(1980){Coleman}, {Wu}, \&
  {Weedman}}]{1980ApJS...43..393C}
{Coleman}, G.~D., {Wu}, C.~C., \& {Weedman}, D.~W. 1980, \apjs, 43, 393

\bibitem[{{Condon}(1974)}]{1974ApJ...188..279C}
{Condon}, J.~J. 1974, \apj, 188, 279

\bibitem[{{de Jong} \& {Lacey}(2000)}]{2000ApJ...545..781D}
{de Jong}, R.~S. \& {Lacey}, C. 2000, \apj, 545, 781

\bibitem[{{de Vaucouleurs}(1953)}]{1953MNRAS.113..134D}
{de Vaucouleurs}, G. 1953, \mnras, 113, 134

\bibitem[{Del~Moral {et~al.}(2012)Del~Moral, Doucet, \& Jasra}]{DelMoral2012}
Del~Moral, P., Doucet, A., \& Jasra, A. 2012, Statistics and Computing, 22,
  1009

\bibitem[{{Drory} {et~al.}(2009){Drory}, {Bundy}, {Leauthaud}, {Scoville},
  {Capak}, {Ilbert}, {Kartaltepe}, {Kneib}, {McCracken}, {Salvato}, {Sanders},
  {Thompson}, \& {Willott}}]{2009ApJ...707.1595D}
{Drory}, N., {Bundy}, K., {Leauthaud}, A., {et~al.} 2009, \apj, 707, 1595

\bibitem[{{Eddington}(1913)}]{1913MNRAS..73..359E}
{Eddington}, A.~S. 1913, \mnras, 73, 359

\bibitem[{{Faber} {et~al.}(2007){Faber}, {Willmer}, {Wolf}, {Koo}, {Weiner},
  {Newman}, {Im}, {Coil}, {Conroy}, {Cooper}, {Davis}, {Finkbeiner}, {Gerke},
  {Gebhardt}, {Groth}, {Guhathakurta}, {Harker}, {Kaiser}, {Kassin},
  {Kleinheinrich}, {Konidaris}, {Kron}, {Lin}, {Luppino}, {Madgwick},
  {Meisenheimer}, {Noeske}, {Phillips}, {Sarajedini}, {Schiavon}, {Simard},
  {Szalay}, {Vogt}, \& {Yan}}]{2007ApJ...665..265F}
{Faber}, S.~M., {Willmer}, C.~N.~A., {Wolf}, C., {et~al.} 2007, \apj, 665, 265

\bibitem[{{Fioc} \& {Rocca-Volmerange}(2019)}]{2019A&A...623A.143F}
{Fioc}, M. \& {Rocca-Volmerange}, B. 2019, \aap, 623, A143

\bibitem[{{Fitzpatrick} \& {Massa}(1990)}]{1990ApJS...72..163F}
{Fitzpatrick}, E.~L. \& {Massa}, D. 1990, \apjs, 72, 163

\bibitem[{{Fitzpatrick} \& {Massa}(2007)}]{Fitzpatricketal_2007}
{Fitzpatrick}, E.~L. \& {Massa}, D. 2007, \apj, 663, 320

\bibitem[{{Fukugita} {et~al.}(1995){Fukugita}, {Shimasaku}, \&
  {Ichikawa}}]{1995PASP..107..945F}
{Fukugita}, M., {Shimasaku}, K., \& {Ichikawa}, T. 1995, \pasp, 107, 945

\bibitem[{{Heavens} {et~al.}(2000){Heavens}, {Jimenez}, \&
  {Lahav}}]{2000MNRAS.317..965H}
{Heavens}, A.~F., {Jimenez}, R., \& {Lahav}, O. 2000, \mnras, 317, 965

\bibitem[{{Hogg} {et~al.}(2002){Hogg}, {Baldry}, {Blanton}, \&
  {Eisenstein}}]{2002astro.ph.10394H}
{Hogg}, D.~W., {Baldry}, I.~K., {Blanton}, M.~R., \& {Eisenstein}, D.~J. 2002,
  arXiv e-prints, astro

\bibitem[{{Hogg} {et~al.}(2003){Hogg}, {Blanton}, {Eisenstein}, {Gunn},
  {Schlegel}, {Zehavi}, {Bahcall}, {Brinkmann}, {Csabai}, {Schneider},
  {Weinberg}, \& {York}}]{2003ApJ...585L...5H}
{Hogg}, D.~W., {Blanton}, M.~R., {Eisenstein}, D.~J., {et~al.} 2003, \apjl,
  585, L5

\bibitem[{{Hudelot} {et~al.}(2012){Hudelot}, {Cuillandre}, {Withington},
  {Goranova}, {McCracken}, {Magnard}, {Mellier}, {Regnault}, {Betoule},
  {Aussel}, {Kavelaars}, {Fernique}, {Bonnarel}, {Ochsenbein}, \&
  {Ilbert}}]{2012yCat.2317....0H}
{Hudelot}, P., {Cuillandre}, J.~C., {Withington}, K., {et~al.} 2012, VizieR
  Online Data Catalog, II/317

\bibitem[{Hwang {et~al.}(1979)Hwang, Masud, \& Paidy}]{rug01:000011811}
Hwang, C.-L., Masud, A. S. M.~M., \& Paidy, S.~R. 1979, Multiple objective
  decision making : methods and applications : a state-of-the-art survey
  (Berlin : Springer)

\bibitem[{{Kacprzak} {et~al.}(2018){Kacprzak}, {Herbel}, {Amara}, \&
  {R{\'e}fr{\'e}gier}}]{2018JCAP...02..042K}
{Kacprzak}, T., {Herbel}, J., {Amara}, A., \& {R{\'e}fr{\'e}gier}, A. 2018,
  \jcap, 2018, 042

\bibitem[{{Kingma} \& {Ba}(2014)}]{2014arXiv1412.6980K}
{Kingma}, D.~P. \& {Ba}, J. 2014, arXiv e-prints, arXiv:1412.6980

\bibitem[{Krizhevsky {et~al.}(2012)Krizhevsky, Sutskever, \&
  Hinton}]{NIPS2012_4824}
Krizhevsky, A., Sutskever, I., \& Hinton, G.~E. 2012, in Advances in Neural
  Information Processing Systems 25, ed. F.~Pereira, C.~J.~C. Burges,
  L.~Bottou, \& K.~Q. Weinberger (Curran Associates, Inc.), 1097--1105

\bibitem[{Lehmann \& Casella(1998)}]{LehmCase98}
Lehmann, E.~L. \& Casella, G. 1998, Theory of Point Estimation, 2nd edn. (New
  York, NY, USA: Springer-Verlag)

\bibitem[{{Lilly} {et~al.}(1995){Lilly}, {Tresse}, {Hammer}, {Crampton}, \& {Le
  Fevre}}]{1995ApJ...455..108L}
{Lilly}, S.~J., {Tresse}, L., {Hammer}, F., {Crampton}, D., \& {Le Fevre}, O.
  1995, \apj, 455, 108

\bibitem[{{L{\'o}pez-Sanjuan} {et~al.}(2017){L{\'o}pez-Sanjuan}, {Tempel},
  {Ben{\'\i}tez}, {Molino}, {Viironen}, {D{\'\i}az-Garc{\'\i}a},
  {Fern{\'a}ndez-Soto}, {Santos}, {Varela}, {Cenarro}, {Moles}, {Arnalte-Mur},
  {Ascaso}, {Montero-Dorta}, {Povi{\'c}}, {Mart{\'\i}nez}, {Nieves-Seoane},
  {Stefanon}, {Hurtado-Gil}, {M{\'a}rquez}, {Perea}, {Aguerri}, {Alfaro},
  {Aparicio-Villegas}, {Broadhurst}, {Cabrera-Ca{\~n}o}, {Castander}, {Cepa},
  {Cervi{\~n}o}, {Crist{\'o}bal-Hornillos}, {Gonz{\'a}lez Delgado}, {Husillos},
  {Infante}, {Masegosa}, {del Olmo}, {Prada}, \&
  {Quintana}}]{2017A&A...599A..62L}
{L{\'o}pez-Sanjuan}, C., {Tempel}, E., {Ben{\'\i}tez}, N., {et~al.} 2017, \aap,
  599, A62

\bibitem[{{Madau} {et~al.}(1996){Madau}, {Ferguson}, {Dickinson}, {Giavalisco},
  {Steidel}, \& {Fruchter}}]{1996MNRAS.283.1388M}
{Madau}, P., {Ferguson}, H.~C., {Dickinson}, M.~E., {et~al.} 1996, \mnras, 283,
  1388

\bibitem[{{Malmquist}(1922)}]{1922MeLuF.100....1M}
{Malmquist}, K.~G. 1922, Meddelanden fran Lunds Astronomiska Observatorium
  Serie I, 100, 1

\bibitem[{{Malmquist}(1925)}]{1925MeLuF.106....1M}
{Malmquist}, K.~G. 1925, Meddelanden fran Lunds Astronomiska Observatorium
  Serie I, 106, 1

\bibitem[{{Marzke}(1998)}]{1998ASSL..231...23M}
{Marzke}, R.~O. 1998, Astrophysics and Space Science Library, Vol. 231, {The
  Galaxy Luminosity Function at Zero Redshift: Constraints on Galaxy
  Formation}, ed. D.~{Hamilton}, 23

\bibitem[{Oh \& Jung(2004)}]{OH20041311}
Oh, K.-S. \& Jung, K. 2004, Pattern Recognition, 37, 1311

\bibitem[{{Pearson} {et~al.}(2014){Pearson}, {Serjeant}, {Oyabu}, {Matsuhara},
  {Wada}, {Goto}, {Takagi}, {Lee}, {Im}, {Ohyama}, {Kim}, \&
  {Murata}}]{2014MNRAS.444..846P}
{Pearson}, C.~P., {Serjeant}, S., {Oyabu}, S., {et~al.} 2014, \mnras, 444, 846

\bibitem[{{Popescu} {et~al.}(2011){Popescu}, {Tuffs}, {Dopita}, {Fischera},
  {Kylafis}, \& {Madore}}]{2011A&A...527A.109P}
{Popescu}, C.~C., {Tuffs}, R.~J., {Dopita}, M.~A., {et~al.} 2011, \aap, 527,
  A109

\bibitem[{{Robin} {et~al.}(2012){Robin}, {Marshall}, {Schultheis}, \&
  {Reyl{\'e}}}]{2012A&A...538A.106R}
{Robin}, A.~C., {Marshall}, D.~J., {Schultheis}, M., \& {Reyl{\'e}}, C. 2012,
  \aap, 538, A106

\bibitem[{{Robin} {et~al.}(2003){Robin}, {Reyl{\'e}}, {Derri{\`e}re}, \&
  {Picaud}}]{2003A&A...409..523R}
{Robin}, A.~C., {Reyl{\'e}}, C., {Derri{\`e}re}, S., \& {Picaud}, S. 2003,
  \aap, 409, 523

\bibitem[{{Robin} {et~al.}(2014){Robin}, {Reyl{\'e}}, {Fliri}, {Czekaj},
  {Robert}, \& {Martins}}]{2014A&A...569A..13R}
{Robin}, A.~C., {Reyl{\'e}}, C., {Fliri}, J., {et~al.} 2014, \aap, 569, A13

\bibitem[{Rubin(1984)}]{10.2307/2240995}
Rubin, D.~B. 1984, The Annals of Statistics, 12, 1151

\bibitem[{{Salimbeni} {et~al.}(2008){Salimbeni}, {Giallongo}, {Menci},
  {Castellano}, {Fontana}, {Grazian}, {Pentericci}, {Trevese}, {Cristiani},
  {Nonino}, \& {Vanzella}}]{2008A&A...477..763S}
{Salimbeni}, S., {Giallongo}, E., {Menci}, N., {et~al.} 2008, \aap, 477, 763

\bibitem[{{Sandage} {et~al.}(1970){Sandage}, {Freeman}, \&
  {Stokes}}]{1970ApJ...160..831S}
{Sandage}, A., {Freeman}, K.~C., \& {Stokes}, N.~R. 1970, \apj, 160, 831

\bibitem[{{Schechter}(1976)}]{1976ApJ...203..297S}
{Schechter}, P. 1976, \apj, 203, 297

\bibitem[{Szegedy {et~al.}(2015)Szegedy, Liu, Jia, Sermanet, Reed, Anguelov,
  Erhan, Vanhoucke, \& Rabinovich}]{Szegedy_2015_CVPR}
Szegedy, C., Liu, W., Jia, Y., {et~al.} 2015, in The IEEE Conference on
  Computer Vision and Pattern Recognition (CVPR)

\bibitem[{{Szegedy} {et~al.}(2015){Szegedy}, {Vanhoucke}, {Ioffe}, {Shlens}, \&
  {Wojna}}]{2015arXiv151200567S}
{Szegedy}, C., {Vanhoucke}, V., {Ioffe}, S., {Shlens}, J., \& {Wojna}, Z. 2015,
  arXiv e-prints, arXiv:1512.00567

\bibitem[{{Taghizadeh-Popp} {et~al.}(2015){Taghizadeh-Popp}, {Fall}, {White},
  \& {Szalay}}]{2015ApJ...801...14T}
{Taghizadeh-Popp}, M., {Fall}, S.~M., {White}, R.~L., \& {Szalay}, A.~S. 2015,
  \apj, 801, 14

\bibitem[{{Tolman}(1934)}]{1934rtc..book.....T}
{Tolman}, R.~C. 1934, {Relativity, Thermodynamics, and Cosmology}

\bibitem[{{Tortorelli} {et~al.}(2020){Tortorelli}, {Fagioli}, {Herbel},
  {Amara}, {Kacprzak}, \& {Refregier}}]{2020arXiv200107727T}
{Tortorelli}, L., {Fagioli}, M., {Herbel}, J., {et~al.} 2020, arXiv e-prints,
  arXiv:2001.07727

\bibitem[{{Trujillo} {et~al.}(2006){Trujillo}, {F{\"o}rster Schreiber},
  {Rudnick}, {Barden}, {Franx}, {Rix}, {Caldwell}, {McIntosh}, {Toft},
  {H{\"a}ussler}, {Zirm}, {van Dokkum}, {Labb{\'e}}, {Moorwood},
  {R{\"o}ttgering}, {van der Wel}, {van der Werf}, \& {van
  Starkenburg}}]{2006ApJ...650...18T}
{Trujillo}, I., {F{\"o}rster Schreiber}, N.~M., {Rudnick}, G., {et~al.} 2006,
  \apj, 650, 18

\bibitem[{{Weyant} {et~al.}(2013){Weyant}, {Schafer}, \&
  {Wood-Vasey}}]{2013ApJ...764..116W}
{Weyant}, A., {Schafer}, C., \& {Wood-Vasey}, W.~M. 2013, \apj, 764, 116

\bibitem[{{Williams} {et~al.}(2010){Williams}, {Quadri}, {Franx}, {van Dokkum},
  {Toft}, {Kriek}, \& {Labb{\'e}}}]{2010ApJ...713..738W}
{Williams}, R.~J., {Quadri}, R.~F., {Franx}, M., {et~al.} 2010, \apj, 713, 738

\bibitem[{{Zucca} {et~al.}(2006){Zucca}, {Ilbert}, {Bardelli}, {Tresse},
  {Zamorani}, {Arnouts}, {Pozzetti}, {Bolzonella}, {McCracken}, {Bottini},
  {Garilli}, {Le Brun}, {Le F{\`e}vre}, {Maccagni}, {Picat}, {Scaramella},
  {Scodeggio}, {Vettolani}, {Zanichelli}, {Adami}, {Arnaboldi}, {Cappi},
  {Charlot}, {Ciliegi}, {Contini}, {Foucaud}, {Franzetti}, {Gavignaud},
  {Guzzo}, {Iovino}, {Marano}, {Marinoni}, {Mazure}, {Meneux}, {Merighi},
  {Paltani}, {Pell{\`o}}, {Pollo}, {Radovich}, {Bondi}, {Bongiorno},
  {Busarello}, {Cucciati}, {Gregorini}, {Lamareille}, {Mathez}, {Mellier},
  {Merluzzi}, {Ripepi}, \& {Rizzo}}]{2006A&A...455..879Z}
{Zucca}, E., {Ilbert}, O., {Bardelli}, S., {et~al.} 2006, \aap, 455, 879

\end{thebibliography}
\bibliographystyle{aa}

\newpage\hbox{}\thispagestyle{empty}\newpage

\appendix

\section{Approximate Bayesian computation (ABC) and population Monte Carlo (PMC)}\label{sec:ABC_PMC}

\subsection{Approximate Bayesian computation (ABC)}\label{sec:ABC}

Because the likelihood function of our model is unknown, we can use approximate Bayesian computation (ABC) to bypass its evaluation by considering the density of forward simulations around some observed data. The first ABC-related ideas date back to the 1980s \citep{10.2307/2240995} with a sampling method that  asymptotically yields the posterior distribution. The ABC-rejection sampling in its most basic form generates $n$ simulations, $\{\hat{d}_i(\theta_i)|i\in [1,n]\},$ from model parameters $\theta_i$ following a prior distribution $p(\theta)$ and compares them to the observed data, $d$. Simulations that are distinctly different from the observed data $d$ are considered unlikely to have been generated from the same parameter values describing $d$, and the associated simulation parameter values are rejected. In more precise terms, the sample $\hat{d_i}$ is accepted with tolerance $\epsilon\geq 0$ if
\begin{equation}
    \rho(\hat{d_i},d)\leq\epsilon
,\end{equation}
where $\rho$ is a distance that measures the discrepancy between $\hat{d_i}$ and $d$. In general, the form of the distance measure, $\rho$, can be difficult to choose for ABC (amounting to a similar interpretation as a choice of likelihood function), but is well motivated when using the IMNN~(\cite{2018PhRvD..97h3004C}). The probability of generating a data set $\hat{d_i}$ with a small distance to $d$ typically decreases as the dimensionality of the data increases. A common approach to somewhat alleviate this problem is to replace $d$ with a set of lower-dimensional summary statistics $S(d)$, which are selected to capture the relevant information in $d$. The full relevance of Sects. \ref{sec:compression} and \ref{sec:NN} for obtaining the summary statistics thus becomes clear in this context, and $S$ is the score compression provided by the IMNN in our application. The acceptance criterion in the ABC rejection algorithm then becomes
\begin{equation}
    \rho(S(\hat{d_i}),S(d))\leq\epsilon
,\end{equation}
where the summary statistics are derived using the compression described in Sect. \ref{sec:compression}. In our application, the distance used between a summary statistic of a simulation $\hat{t_i}=S(d_i)$ and that of the observed data, $t_\text{obs}=S(d)$, is defined by the Fisher information of the previously trained network \citep{2018PhRvD..97h3004C} by
\begin{equation}
    \rho(\hat{t_i},t_\text{obs} )=\left(\widetilde{\theta}(\hat{t_i})-\widetilde{\theta}(t_\text{obs})\right)^TF\left(\widetilde{\theta}(\hat{t_i})-\widetilde{\theta}(t_\text{obs})\right)\label{eq:dist}
.\end{equation}
This distance measure is justified with the quasi-maximum likelihood estimates $\widetilde{\theta}$ described in Eq. \eqref{eq:MLE_summaries} because that space should be asymptotically Euclidean (with the Fisher information as the metric) once the network has converged and the summary statistics become Gaussian distributed \citep{2018PhRvD..97h3004C}.

Even when using summary statistics, the acceptance rate of the ABC is low for a small $\epsilon$ because we sample from the whole multidimensional prior distribution. This means that finding a new set of parameters whose simulations are more similar to the data is unlikely, which makes the sampling from the prior distribution inefficient and hence slow. Therefore we describe the faster population Monte Carlo (PMC) procedure in the next section.

\subsection{Population Monte Carlo (PMC)}\label{sec:PMC}

The goal of the PMC procedure is to have a higher density of draws in the more probable regions of the posterior distribution. The different steps of the PMC used here are listed below.
\begin{itemize}
    \item \textbf{Step 1:} From the prior distribution, we draw $N$ (user-defined) sets of parameters. For each set of parameters in this sample, we follow the steps listed below.
    \begin{itemize}
        \item[$\star$] \textbf{Step 1a:} We simulate the multiband images.
        \item[$\star$] \textbf{Step 1b:} We compress the simulated images with IMNN to reduce the dimensional complexity.
        \item[$\star$] \textbf{Step 1c:} We compute the distance between the summarized simulation and the summary of the observed data using Eq. \eqref{eq:dist}.
        \item[$\star$] \textbf{Step 1d:} We define a weighting that corresponds to the probability of this set of parameters under the prior distribution.
    \end{itemize}
    \item \textbf{Step 2:} We compute the mean vector and the weighted-covariance matrix of the $N$ sets of parameters. We set the $R$ counter to $0$.
    \item \textbf{Step 3:} From the $N$ sets of\ parameters, we identify the $Q\%$ (user-defined) of the compressed simulations that have the largest distance from the summary of the observed data. For each set of parameters in this subsample, we follow the steps listed below.
    \begin{itemize}
        \item[$\star$] \textbf{Step 3a:} We resample it from a proposal distribution that is a Gaussian using the current parameter values as the mean and the weighted-covariance matrix from Step 2. Each time we go through this step, we increment the $R$ counter.
        \item[$\star$] \textbf{Step 3b:} We simulate the multiband images at this new proposed set of parameters.
        \item[$\star$] \textbf{Step 3c:} We compress the simulated image with IMNN to reduce the dimensional complexity.
        \item[$\star$] \textbf{Step 3d:} We compute the distance between the summarized simulation and the summary of the observed data using Eq. \eqref{eq:dist}.
        \item[$\star$] \textbf{Step 3e:} If this distance is smaller than the previous distance, we keep this set of parameters. Otherwise, we return to \textbf{Step 3a}.
    \end{itemize}
    \item \textbf{Step 4:} Each weight is updated using the probability of the corresponding resampled set of calculated parameters and the weighted-covariance matrix of Step 2. Each initial weight (under the prior distribution) is divided by the normalized Gaussian using the difference between the new and initial parameter set as the mean and the weighted-covariance matrix from Step 2.
    \item \textbf{Step 5:} As long as the counter $R$ is smaller than the user-defined threshold $M$, we return to \textbf{Step 2} and re-identify the $Q\%$ simulations with the largest distance from the observed data. Otherwise, the PMC concludes. 
\end{itemize}
    
The $Q\%$ in Step 3 is a user choice, and we chose values allowing us to somewhat parallelize the procedure. If the number of draws $M$ is large enough, the distribution of the $N$ sets of parameters has become approximately stationary when the procedure stops, and it can then be considered as a good approximation of the posterior \citep{DelMoral2012}.

In the application of Sect. \ref{sec:app1}, we used $Q=75\%$ and $N=1000,$ and we adopted $M=50000$. At each iteration, the procedure therefore tries to re-draw $750$ samples and concludes when the number of attempts in the same iteration is higher than $50000$.

In the application of Sect. \ref{sec:app2}, we used $Q=25\%$ and $N=500,$ and we adopted $M=5000$. At each iteration, the procedure therefore tries to re-draw $125$ samples and concludes when the number of attempts in the same iteration is higher than $5000$.

\end{document}